\documentclass[11pt]{article} 
\newcommand{\version}{February, 2008}

\usepackage{amsthm,amsfonts,amsmath,amssymb}
\usepackage{graphicx}

\swapnumbers
\theoremstyle{plain}

\newtheorem{thm}{THEOREM}[section]
\newtheorem{lm}[thm]{LEMMA}

\newtheorem{prop}[thm]{PROPOSITION}
\newtheorem{remark}[thm]{Remark}

\theoremstyle{definition}

\theoremstyle{remark}
\newcommand{\upchi}{\raise1pt\hbox{$\chi$}}
\newcommand{\R}{{\mathord{\mathbb R}}}

\newcommand{\dd}{{\rm d}}

\renewcommand{\|}{{\Vert}}
\def\gg{\widehat{g}}
\def\hh{{\widehat{h}}}

\def\ee{{\rm e}}

\numberwithin{equation}{section}  

\begin{document}

%%%%%%%%%%%%%%%%%%%%%%
%\markboth{\scriptsize{CCC \version}}{\scriptsize{CCC \version}}

\title{Strong Convergence towards homogeneous cooling states for dissipative Maxwell models}
%\title{UNIFORM PROPAGATION OF REGULARITY FOR DISSIPATIVE KINETIC EQUATIONS}
\author{
\vspace{5pt}  Eric A. Carlen$^{1}$  Jos\'e A. Carrillo$^{2}$ and Maria C. Carvalho$^{3}$,\\
\vspace{5pt}\small{$1.$ Department of Mathematics, Hill Center,}\\[-6pt]
\small{Rutgers University,
110 Frelinghuysen Road
Piscataway NJ 08854-8019 USA}\\
\vspace{5pt}\small{$2.$ Instituci\'o Catalana de Recerca i Estudis Avan\c cats and Departament de Matem\`atiques}\\[-6pt]
\vspace{5pt}\small{Universitat Aut\`onoma de Barcelona, E-08193 Bellaterra, Spain}\\[-6pt]
\vspace{5pt}\small{$3.$ CMAF and Departamento de Matematica da Faculdade de Ci\^encias da Universidade de Lisboa}\\[-6pt]
\vspace{5pt}\small{1640--003 Lisboa, Portugal}\\[-6pt]
 }
\date{\version}
\maketitle

\footnotetext[4]{EC is partially supported by U.S. National
Science Foundation grant DMS 06-00037 and PHY 01-39984. JAC
partially supported by the DGI-MEC (Spain) FEDER-project
MTM2005-08024. MCC is partially supported by FCT POCI/MAT/61931.
\\
\copyright\, 2008 by the authors. This paper may be reproduced, in
its entirety, for non-commercial purposes.}

\begin{abstract}
We show the propagation of regularity, uniformly in time, for the
scaled solutions of the inelastic Maxwell model for small
inelasticity. This result together with the weak convergence
towards the homogenous cooling state present in the literature
implies the strong convergence in Sobolev norms and in the $L^1$
norm towards it depending on the regularity of the initial data.
The strategy of the proof is based on a precise control of the
growth of the Fisher information for the inelastic Boltzmann
equation. Moreover, as an application we obtain a bound in the
$L^1$ distance between the homogeneous cooling state and the
corresponding Maxwellian distribution vanishing as the
inelasticity goes to zero.
\end{abstract}

\bigskip
\centerline{Mathematics Subject Classification Numbers: 82C40,
35B40}

\bigskip
\section{Introduction} \label{MS}
\medskip

This paper concerns the regularity properties of solutions of the
spatially homogeneous Boltzmann equation for Maxwellian molecules
in $\R^3$ with inelastic collisions, introduced in
\cite{Bobylev-Carrillo-Gamba}. This equation describes the
evolution of the distribution of the velocities in a collection of
particles as they interact through inelastic  binary collisions.
Let $f(v,t)$ be the probability density for the velocity of a
particle chosen randomly from the collection at time $t$. Let
$\varphi$ be any bounded and continuous function on $\R^3$. Then
the equation under investigation is given, in weak form, by
\begin{equation}\label{lis1}
\frac{{\rm d}}{{\rm d}t}\langle f, \varphi\rangle =  \langle
Q_e(f,f), \varphi\rangle
\end{equation}
where $\langle f, \varphi\rangle$ denotes
$\int_{\R^3}f(v,t)\varphi(v)\dd v$, and where
\begin{equation}\label{lis2}
 \langle Q_e(f,f), \varphi\rangle = \frac{1}{4\pi}\int_{\R^3}\int_{\R^3}\int_{S^2}
 \widetilde B\left(n\cdot \frac{v-w}{|v-w|}\right)f(v)f(w)[\varphi(v') - \varphi(v)]\,\dd n\,\dd v\,\dd w
\end{equation}
where $n$ is a unit vector in $S^2$, $\dd n$ is the uniform
measure on $S^2$ with total mass $4\pi$, with $v'$ the post
collisional velocity given by
\begin{equation}\label{lis3}
 v'  =  v -\frac{1+e}{2}((v-w)\cdot n) n\ ,
\end{equation}
with $0\le  e \le 1$, and where $\widetilde B$ is a positive,
integrable even function on $[-1,1]$. Because of the integrability
of $\widetilde B$, we can separate the collision operator in the
gain and loss terms, $Q_e(f,f):=Q^+_e(f,f)-Q^-_e(f,f)$ with
$$
Q^-_e(f,f):= \left(\int_{S^2} \widetilde B\left(n\cdot
\frac{v-w}{|v-w|}\right)\,\dd n\right) f = \left( \frac12
\int_{-1}^1 \widetilde B(s) \, ds \right) f.
$$
The function $\widetilde B$ gives the rate at which the various
kinematically possible collisions happen, and the tilde is present
because later on we shall also consider another rate function $B$
corresponding to another parameterization of the kinematically
possible collisions.

The parameter $e$ is the {\em restitution coefficient}. For $e <
1$, the collisions are inelastic, and energy is dissipated in each
collision. In this case, the collisions are not reversible. This
is a crucial difference with the elastic theory in which there is
a complete time reversal  symmetry between the pre and post
collisional velocities. It is for this reason that we have written
(\ref{lis1}) in weak form, and not because of any difficulty in
constructing strong solutions: It is just that to write the
equation down in strong form, we need a parameterization of the
possible pairs of precollisional velocities $(v^*,w^*)$ that can
result in the pair of post collisional velocities $(v,w)$. We have
detailed the strong formulation in the different parameterizations
in the Appendix, see \cite{Bobylev-Carrillo-Gamba,GPV} for
particular cases, but for the purposes of the introduction, the
weak form specifies the equation under consideration well enough
for us to proceed with the description of the particular issues
with which we are concerned, and the results we obtain.

The reason we require $\widetilde B$ to be even is that the post
collisional velocity $v'$ defined in (\ref{lis3}) depends on $n$
quadratically, and thus is unchanged under the substitution $n\to
-n$. {\em For this reason, we may freely assume $\widetilde B$ to
be even, and we do so in what follows.}

The first thing to notice about the equation is that the first
moment of $f$ is conserved. Indeed, for any $a\in \R^3$, let
$\varphi(v) = a\cdot v$. Then we have from (\ref{lis1}) and
(\ref{lis3})  that
\begin{eqnarray*}
\frac{{\rm d}}{{\rm d}t}\langle f, \varphi\rangle &=&
-\frac{1+e}{8\pi}\int_{\R^3}\int_{\R^3}\int_{S^2}\widetilde
B\left(n\cdot \frac{v-w}{|v-w|}\right)f(v)f(w)
[((v-w)\cdot n)(a\cdot n)]\dd v\dd w\dd n\nonumber\\
&=& -\frac{1+e}{4}\left[\int_{-1}^{1} s^2 \widetilde B(s)\dd
s\right]\int_{\R^3}\int_{\R^3}f(v)f(w)a\cdot(v-w)\dd v\dd w = 0\ .
\end{eqnarray*}
Indeed, as detailed in the appendix, the companion formula to
(\ref{lis3}), giving the other post collisional velocity $w'$, is
\begin{equation*}
 w'  =  w +\frac{1+e}{2}((v-w)\cdot n) n\ .
\end{equation*}
Thus, in each individual collision $(v,w) \to (v',w')$, the total
momentum $v+w$ is conserved, and this certainly ensures that the
first moment of $f$ is conserved.  In any case, on account of the
computation just made, we may as well assume that our initial data
$f_0$ satisfies
\begin{equation*}\label{lis6a}
  \int_{\R^3}vf_0(v)\,\dd v = 0\ .
\end{equation*}
Then of course we shall have
\begin{equation}\label{lis6}
 \int_{\R^3}vf(v,t)\,\dd v = 0
\end{equation}
for all $t\geq 0$. While momentum is conserved,  energy is
dissipated, as we have indicated above.  We now calculate the rate
of this dissipation: Take $\varphi(v) = |v|^2$ and then note that
from (\ref{lis3})
$$
\varphi(v') = |v|^2 - (1+e)((v-w)\cdot n)(v\cdot n) + \frac{(1+e)^2}{4}((v-w)\cdot n)^2\ .
$$
In this case, using the abreviated notation $u = v-w$, we have
\begin{eqnarray*}
\frac{{\rm d}}{{\rm d}t}\langle f, \varphi\rangle \!\!&=&
\!\!\frac1{4\pi} \int_{\R^3}\!\int_{\R^3}\!\int_{S^2}\!\widetilde
B\left(n\cdot \frac{u}{|u|}\right)f(v)f(w)
\left[ \frac{(1+e)^2}{4}(u\cdot n)^2  - (1+e)(u\cdot n)(v\cdot n) \right]\dd v\dd w\dd n\nonumber\\
&=& \left[\frac12 \int_{-1}^{1}s^2  \widetilde B(s)\dd s\right]\int_{\R^3}\int_{\R^3}f(v)f(w)
\left(\frac{(1+e)^2}{4}|u|^2 - (1+e)u\cdot v\right)\dd v\dd w \nonumber\\
&=& -\left[\frac12 \int_{-1}^{1}s^2  \widetilde B(s)\dd
s\right]\frac{1-e^2}{2}\langle f, \varphi\rangle\ .
\end{eqnarray*}
That is, with the positive constant $E$ defined by
\begin{equation}\label{Edef}
E = \left[\frac12\int_{-1}^{1}s^2  \widetilde B(s)\dd
s\right]\frac{1-e^2}{4}\ ,
\end{equation}
we have every solution of (\ref{lis1}) with initial data $f_0$
satisfying (\ref{lis6a}) satisfies
\begin{equation}\label{lis6c}
 \frac{{\rm d}}{{\rm d}t}\int_{\R^3}|v|^2f(v,t)\dd v = -2E \int_{\R^3}|v|^2f(v,t)\dd v\ .
\end{equation}

This implies that  $f(v,t)\dd v$ tends to a point mass at $v=0$ as
$t$ tends to infinity. It is natural to enquire into precise
nature of this collapse to a point mass. In previous works
\cite{Bobylev-Cercignani2,Bobylev-Cercignani-Toscani,BisiCT2,Bobylev-Cercignani-Gamba,Bobylev-Cercignani-Gamba2,Bolley-Carrillo},
it has been shown that if one rescales $f(v,t)$ to keep the
variance (i.e., temperature) constant, then the rescaled density
tends to a particular equilibrium state, known as the {\em
homogeneous cooling state}. That is, if we define the probability
density $g(v,t)$ by
  \begin{equation}\label{gdef}
g(v,t) = \ee^{-3Et}f\left(\ee^{-Et}v,t\right)\ ,
 \end{equation}
 $$\int_{\R^3}|v|^2g(v,t)\dd v = \int_{\R^3}|v|^2g(v,0)\dd v =   \int_{\R^3}|v|^2 f_0(v)\dd v$$
 for all $t$, and
 there is a density $g_\infty$ such that
   \begin{equation}\label{hcsdef}
\lim_{t\to\infty}g(v,t) = g_\infty(v)\ .
 \end{equation}
The convergence in (\ref{hcsdef}), part of the so-called
Ernst-Brito conjecture
\cite{EB1,EB2,Bobylev-Cercignani2,Bobylev-Cercignani-Toscani}, has
so far been shown in certain weak norms that we shall introduce
shortly, see \cite{CT} for a review and
\cite{Bobylev-Cercignani-Toscani,BisiCT2} for the proofs. Our goal
in this paper is to prove that $g(v,t)$ is regular in $v$,
uniformly in $t$. This is reasonable to expect since it is was
proved by Bobylev and Cercignani that $g_\infty(v)$ itself is
quite regular \cite[Theorem 7.1]{Bobylev-Cercignani2}. However, it
is clear from the fact that $f(v,t)\dd v$ tends to a point mass
that the norm of $f(\cdot,t)$ must diverge with $t$ in every norm
that would imply smoothness of $f(\cdot,t)$. While the rescaling
may well lower such norms, one needs very precise estimates on the
loss of regularity to avoid having them overwhelm whatever one
gains from the rescaling. Notice in particular that the rescaling
does nothing to improve regularity.

To investigate the long time behavior of $g(v,t)$, we write down
its evolution equation, which of course is obtained from
(\ref{lis1}) through (\ref{gdef}). In working out the equation, we
make use of the dilation invariance of the collision  integral
$Q_e(f,f)$:  For any density $f$, test function $\varphi$ and any
$\lambda>0$, define
    \begin{equation}\label{scaling}
    f^{(\lambda)}(v) = \lambda^3f(\lambda v)\qquad{\rm and}\qquad  \varphi^{(\lambda)}(v) = \varphi(v/\lambda)\ .
    \end{equation}
Then one easily sees from (\ref{lis2}) that
    \begin{equation}\label{lis9}
    \langle Q_e(f^{(\lambda)},f^{(\lambda)}),\varphi\rangle =  \langle Q_e(f,f),\varphi^{(\lambda)}\rangle\ .
    \end{equation}
Then, for any test function $\varphi$,
\begin{eqnarray*}%\label{gform}
 \frac{{\rm d}}{{\rm d}t}
 \langle g,\varphi\rangle &=&   \frac{{\rm d}}{{\rm d}t}
 \langle f,\varphi^{(\exp(-Et))}\rangle\nonumber\\
 &=&-E\ee^{-Et}\langle f,v\cdot\left(\nabla \varphi\right)^{(\exp(-Et))}\rangle + \langle Q( f,f),\varphi^{(\exp(-Et))}\rangle\nonumber\\
 &=&-E\langle g,\left(v\cdot\nabla \varphi\right)\rangle + \langle Q( f,f),\varphi^{(\exp(-Et))}\rangle\nonumber\\
 &=&-E\langle g,\left(v\cdot\nabla \varphi\right)\rangle + \langle Q( g,g),\varphi\rangle
\end{eqnarray*}
where we have used (\ref{lis9}) in the last line. Thus, our
evolution equation for $g$ is, in weak form,
  \begin{equation}\label{lis10}
   \frac{{\rm d}}{{\rm d}t}
 \langle g,\varphi\rangle =  -E\langle g,\left(v\cdot\nabla \varphi\right)\rangle + \langle Q( g,g),\varphi\rangle\ .
    \end{equation}

There are other ways, physically and mathematically different, to
control the temperature/variance: If the particles are in contact
with an appropriate heat bath, this will add a thermal
regularization to the evolution equation for $f$. This thermal
bath can be modelled by stochastic heating, i.e.,
$$
\frac{\partial f}{\partial t} = Q_e(f,f) + \Delta_v f
$$
or by a thermalized bath of particles, adding a linear Boltzmann
type operator. In these two cases, global regularity estimates for
solutions have been obtained, see \cite{BisiCT,CT,Z06}. However,
as the first order anti--drift term in (\ref{lis10}) does not
induce a priori any regularization, the problem of proving global
regularity estimate for solutions of (\ref{lis10}) is more
challenging.

The {\em Fisher information} plays a crucial role in our
investigation of regularity. For any probability density $f$ on
$\R^3$, the Fisher information, $I(f)$, is defined by
$$
I(f)  = 4\int_{\R^3}|\nabla \sqrt{f(v)}|^2\,\dd v =
\int_{\R^3}|\nabla \ln f(v)|^2f(v)\,\dd v
$$
whenever the distributional gradient of $\sqrt{f}$ is square
integrable, and it is defined to be infinite otherwise. It has
been shown by Villani that in case $e=1$; i.e, for elastic
collisions, the Fisher information is non increasing in time. This
is a basic  propagation of regularity result that is the starting
point of our investigation of the inelastic case.

For solutions of (\ref{lis10}), the Fisher information will not be
bounded uniformly in time.  Indeed, the Fisher information has
simple scaling properties: If $f^{(\lambda)}$ is defined in terms
of $f$ and $\lambda$ as in (\ref{scaling}), one easily computes
   \begin{equation}\label{fishscale}
   I(f^{(\lambda)}) = \lambda^2 I(f)\ .
   \end{equation}
Therefore, with $g(v,t)$ defined in terms of $f(v,t)$ through the
scaling relation (\ref{gdef}), we have
   \begin{equation}\label{fgish}
   I(g(\cdot,t)) = \ee^{-2Et}I(f(\cdot,t))\ .
   \end{equation}
The exponentially decreasing factor $\ee^{-2Et}$ is good, but
notice from (\ref{Edef}) that it depends only on $\widetilde B$
and the restitution coefficient $e$, and not on the initial data.
For some initial data, $I(f(\cdot,t))$ will grow faster than this
rate, and thus $ I(g(\cdot,t)) $ will grow exponentially.
Nonetheless, we shall be able to prove that its growth is not too
bad, at least for $e$ not too far from $1$.

\begin{thm}\label{fishpro}
For any solution $g(v,t)$ of \eqref{lis10}, we have a bound on the
Fisher information
$$
I(g(\cdot, t)) \le \ee^{[(1-e)(2+e+15e^2)/(8e^3) - 2E]t}
I(g(\cdot, 0))
$$
where $E$ is the constant defined in \eqref{Edef}. While the
exponent is always positive, no matter how $\widetilde B$ is
chosen, it does always vanish in the limit $e\to 1$.
\end{thm}

Theorem \ref{fishpro} is proved in Section 2. Our main goal in the
next part of the paper is to obtain a tiny uniform-in-time
propagation of regularity result of the type:

\begin{thm}\label{mainintro}
For any $0<\delta<1$, there is a computable positive constant $C$,
such that for any solution $g$ of \eqref{lis10} corresponding to
the initial value~$f_0$ with unit mass, zero mean velocity,
$|v|^{2+\alpha}f_0\in L^1 (\R^3)$ with $0<\alpha<1$ and $I(f_0) <
\infty$, then
\begin{equation}\label{new}
\||\eta|^{\delta} \hat{g}(\eta)\|_{L^\infty(\R^3)} \le C,
\end{equation}
for all $t>0$, being $e$ close enough to 1.
\end{thm}

This strategy precisely coincides with the open problem left in
\cite{CT} for strong convergence to homogeneous cooling states and
applied in the case of the thermalized bath of particles, adding a
linear Boltzmann type operator, see \cite[Subsubsection
7.2.4]{CT}.

To prove the convergence in strong norms towards the homogeneous
cooling state, we will need more; we will need the propagation of
regularity in Sobolev spaces of high degree:
\begin{equation}\label{Sob_n}
 \| g \|_{\dot{H}^r(\R^3)}^2= \int_{\R^3} |\eta|^{2r}|\hat
g(\eta)|^2\, d\eta
\end{equation}
with $r>0$. However, there is a well-developed machinery
\cite{CGT,CT} for showing that whenever the equation propagates a
tiny degree of regularity, this implies the equation propagates
regularity of any degree. Therefore, the main problem to be solved
is to prove \eqref{new}, uniformly in time for which there are no
standard arguments.

Then, using the regularity in high Sobolev spaces, we can parley
the weak convergence in (\ref{hcsdef}) into convergence in all
Sobolev norms, Theorem \ref{hrconv}, and strong $L^1$ convergence
at an explicit exponential rate for a certain class of initial
data. This is the objective of Section 3 and the main result is
summarized as:

\begin{thm}\label{l1conv}
Given the solution $g$ of \eqref{lis10} corresponding to the
initial probability distribution function~$f_0\in
\dot{H}^r(\R^3)$, with $r>0$, of zero mean velocity such that
$|v|^4 f_0\in L^1 (\R^3)$ and $I(f_0) < \infty$. Then, for $e$
close to 1, the solution~$g(t,v)\!$ of~\eqref{lis10} converges
strongly in $L^1$ with an exponential rate towards the homogenous
cooling state, i.e., there exist positive constants $C$ and
$\gamma'$ explicitly computable such that
$$
\|g(t)-g_\infty \|_{L^1(\R^3)} \leq C\, \ee^{-\gamma' t}
$$
for all $t>0$.
\end{thm}

Finally, we can study the small inelasticity limit of the sequence
of homogeneous cooling states showing an $L^1$ convergence towards
the Maxwellian distribution with zero mean velocity and
temperature fixed by the initial data as $e\to 1$ with an explicit
speed in terms of the inelasticity parameter. Section 4 is devoted
to this small inelasticity limit in strong norms. Finally, as
announced above, the appendix is aimed at a detailed description
of the relations between the different parameterizations of the
collision mechanism that we have written for non necessarily
Maxwellian type collision kernels.

%%%%%%%%%%%%%%%%%%%%%%%%%%%%%%%%%%%%%%%%%%%%%%%%%%%%%%%%%%%%%%%%%%%%

\section{Fisher Information bounds}\label{fishinf}

Villani \cite{Vil98} has proved that for Maxwellian molecules and
elastic collisions, the Fisher information does not increase. A
special case of this, namely with $\widetilde B$ constant, had
been treated earlier by Carlen and Carvalho using the reflection
parameterization \cite{CC}. Villani's analysis is based on the
$\sigma$ representation, and has the advantage it allows an
arbitrary rate function $B$. All of these results use the strong
formulation of the collision operator, the passage from the weak
form to the strong form is merely a complicated change of
variables that we detail in the appendix. The main formulas we
will use in this section are related to the strong formulation in
the $\sigma$-representation,
$$
Q_e^+(f,f)(v)   = \frac{1}{4\pi} \int_{\R^3}\int_{S^2} f(v^*)
f(w^*) B_e^+ (k\cdot \sigma)\dd \sigma \dd w
$$
with $u=v-w$, $k = u/|u|$,
$$
B_e^+(s) = B\left(\frac{(1+e^2)s - (1-e^2)}{(1+e^2) -
(1-e^2)s}\right) \frac{\sqrt{2}}{\sqrt{{(1+e^2) -
(1-e^2)s}}}\frac{1}{e} \qquad , \qquad B(s) = \frac{
 \widetilde B\left(\sqrt{(1-s)/2}\right)}{2\sqrt{(1-s)/2}}
$$
and the precollisional velocities are given by
$$
\begin{cases}  v^*  &=\  {\displaystyle \frac{v+w}{2} - \frac{1-e}{4e}(v-w) +
\frac{1+e}{4e}|v-w|\sigma}\\[3mm]
 w^*  &=\     {\displaystyle\frac{v+w}{2} + \frac{1-e}{4e}(v-w) - \frac{1+e}{4e}|v-w|\sigma
 }
\end{cases} \, .
$$
The reader can understand now why we have avoided the strong
formulation as long as we could. We emphasize that this operator
coincides with the one defined in weak form below \eqref{lis2}.
Full details of the passage from one representation to the other
are given in the appendix.

We now start to adapt Villani's analysis to the inelastic case,
and derive bounds on the growth of the Fisher information in terms
of the restitution coefficient $e$. The crucial feature of these
bounds on the growth is that they vanish as $e$ tends towards $1$.
The main result of this section is:

\begin{thm}\label{fgrow} For all probability densities $f$ on $\R^3$,
\begin{equation}\label{lis31}
I(Q_e^+(f,f)) \le \left[ 1+
(1-e)\left(\frac{2+e+15e^2}{8e^3}\right)\right]\,I(f)\ ,
\end{equation}
with the consequence that if $f(v,t)$ is a solution of
\eqref{lis1}, we have
\begin{equation}\label{lis32}
I(f(\cdot, t)) \le \ee^{[(1-e)(2+e+15e^2)/(8e^3)]t} I(f(\cdot,
0))\ .
\end{equation}
\end{thm}

As an immediate consequence of this, we obtain the proof of
Theorem \ref{fishpro}:
\medskip

\noindent{\bf Proof of Theorem \ref{fishpro}:} Consider any
rescaled solution $g(v,t)$; i.e., any solution of (\ref{lis10}).
 By (\ref{fgish}), any solution $g(v,t)$ of (\ref{lis10}) satisfies
\begin{equation}\label{lis33}
I(g(\cdot, t)) \le \ee^{[(1-e)(2+e+15e^2)/(8e^3) - 2E]t}
I(g(\cdot, 0))\ .
\end{equation}
where $E$ is given by (\ref{Edef}). Notice that while $E$ depends
on the particular choice of $\widetilde B$, for any choice we have
$$
E \le (1-e)\frac{1+e}{4} < \frac{1-e}{2}\ .
$$
Therefore, for any $\widetilde B$, the exponent in (\ref{lis33})
is at least
$$
(1-e)\left(\frac{2+e+15e^2}{8e^3} -1\right) > 0\ ,
$$
for all $0\le e <1$.\qed

\

While the exponent in Theorem \ref{fishpro} is always positive, it
does vanish in the elastic limit, and that is what we shall need
in the next sections. We begin by recalling several results:

\medskip
\begin{lm}\label{vill1} {\rm \cite[Lemma 1]{Vil98}} Let $P_k$ denote the orthogonal projection onto the span of $k$,
and let $P^\perp_k$ denote its orthogonal complement. Then for any
differentiable rate function $B$,
$$
\nabla_v\left[ B(k\cdot \sigma)\right] = \frac{1}{|u|} B'(k\cdot\sigma)P^\perp_k\sigma\ ,
$$
where $B'$ is the derivative of $B$.
\end{lm}

The proof of this Lemma is an elementary computation. Applying it
with $B = B_e^+$, and defining $F(v,w,\sigma)=f(v^*)f(w^*)$ we
have
\begin{equation}\label{bonn2}
\nabla_v Q_e^+(f,f) = \frac{1}{4\pi}\int_{S^2}\!\int_{\R^3}\!\!
\left[ \frac{1}{|u|}(B_e^+)'(k\cdot \sigma) P_k^\perp \sigma
F(v,w,\sigma) + B_e^+(k\cdot \sigma)\nabla_v
F(v,w,\sigma)\right]\dd w\dd \sigma\ .
\end{equation}
The proof of the next Lemma is not so elementary, as done in
\cite{Vil98}, it is an ingenious integration by parts on the
sphere. Later on, we will give a different proof of the formula
resulting from this Lemma in a more direct way.

\medskip
\begin{lm}\label{vill2}{\rm \cite[Lemma 2]{Vil98}} Using the notation of the previous lemma and also defining the linear transformation $M_{\sigma,k}$
on $\R^3$ by
$$M_{\sigma,k}(x) = (k\cdot \sigma)x - (k\cdot x)\sigma\ ,$$
we have that for any smooth function $F(\sigma)$ on $S^2$,
$$\int_{S^2} B'(k\cdot\sigma)P^\perp_k\sigma F(\sigma) \dd \sigma = \int_{S^2} B(\sigma\cdot k) M_{\sigma,k}\left[
\nabla_\sigma F(\sigma)\right] \dd \sigma\ .
$$
\end{lm}

Using this Lemma on (\ref{bonn2}) we obtain
\begin{equation}\label{bonn3}
\nabla_v Q_e^+(f,f) = \frac{1}{4\pi}\int_{S^2}\!\int_{\R^3}\!\!
\left[ \frac{1}{|u|}B_e^+(\sigma\cdot k) M_{\sigma,k}
\left[\nabla_\sigma F(v,w,\sigma)\right] + B_e^+(k\cdot
\sigma)\nabla_v F(v,w,\sigma)\right]\dd w\dd \sigma\ .
\end{equation}
Next, using (\ref{inqnv}) to evaluate the Jacobians,
$$
\nabla_v\left[ f(v^*)\right] = \left(\frac{\partial v^*}{\partial
v}\right) (\nabla_v f)(v^*) = \frac{3e-1}{4e} (\nabla_v f)(v^*)
+ \frac{1+e}{4e} (\sigma \cdot \nabla_v f)(v^*) k
$$
and
$$
\nabla_v\left[ f(w^*)\right] = \left(\frac{\partial w^*}{\partial
v}\right) (\nabla_v f)(w^*) = \frac{e+1}{4e} (\nabla_v f)(w^*)   -
\frac{1+e}{4e}(\sigma \cdot \nabla_v f)(w^*) k \, .
$$
Also from (\ref{inqnv}),
$$
\nabla_\sigma F(v,w,\sigma) = \frac{1+e}{4e}|u| \left[  (\nabla_v f)(v^*) f(w^*) -   f(v^*)(\nabla_v f)(w^*) \right]\ .
$$
Therefore, if we define the linear transformation $P_{\sigma,k}$
on $\R^3$ by $P_{\sigma,k}(x) = (\sigma\cdot x)k +
M_{\sigma,k}(x)$, we can rewrite (\ref{bonn3}) as
\begin{equation}\label{bonn4}
\nabla_v Q_e^+(f,f) = \frac{1}{4\pi}\int_{S^2}\int_{\R^3}
B_e^+(\sigma\cdot k) G(v,w,\sigma)\dd\sigma\dd w
\end{equation}
where
\begin{align}
G(v,w,\sigma) =&\,\,\,\,\,\,\,\,
f(w^*)\left(\frac{3e-1}{4e} + \frac{1+e}{4e}P_{\sigma,k}\phantom{.}\right)\left[(\nabla_v f)(v^*)\right]\nonumber\\
&+ f(v^*\phantom{.})\left(\frac{1+e}{4e}\ - \
\frac{1+e}{4e}P_{\sigma,k}\right)\left[(\nabla_v f)(w^*)\right] \
. \label{bonn5}
\end{align}

Before proceeding further, let us give a simple, direct proof of
formula \eqref{bonn5} making use of the Fourier transform instead
of Lemma \ref{vill2}.

\

\noindent{\bf Proof of \eqref{bonn4}-\eqref{bonn5}:} We start by
recalling the formula of the Fourier representation of
$Q_e^+(f,f)$ obtained in \cite{Bobylev-Carrillo-Gamba}, see
previous works \cite{Boby75,Bobylev} and \cite{Desvillettes,CT}
for a review. It holds
\begin{equation}\label{fourier}
\widehat{Q_e^+(f,f)}(\eta) = \frac{1}{4 \pi} \int_{S^2}
B(\tilde{\eta}\cdot \sigma) \hat{f} (t, \eta_-) \hat{f} (t,
\eta_+) \, d\sigma
\end{equation}
with $\tilde{\eta}=\eta/|\eta|$ and
\begin{equation}
\begin{array}{l}
\displaystyle \eta_-=\frac{1+e}{4} (\eta-|\eta|\sigma), \vspace{0.3 cm}\\
\displaystyle \eta_+=\frac{3-e}{4}\, \eta + \frac{1+e}{4}\,
|\eta|\sigma = \eta-\eta_-\,.
\end{array}
\label{k+-}
\end{equation}
Now, let us point out the following identity, left for the reader
to check,
$$
Z(\eta_+,\eta_-):=\frac{3e-1}{4e} \eta_+ +
\frac{1+e}{4e}P_{\sigma,\tilde{\eta}}(\eta_+)+ \frac{1+e}{4e}
\eta_- - \ \frac{1+e}{4e}P_{\sigma,\tilde{\eta}}(\eta_-) = \eta +
\frac{1-e^2}{4e}\left( (\eta\cdot\sigma)\tilde{\eta} -
|\eta|\sigma\right).
$$
Now, multiplying both sides by
$\frac{1}{4 \pi} B(\tilde{\eta}\cdot \sigma) \hat{f} (t, \eta_-)
\hat{f} (t, \eta_+)$ and integrating over the sphere, we get
\begin{equation}\label{newfourier}
\frac{1}{4 \pi} \int_{S^2} B(\tilde{\eta}\cdot \sigma) \hat{f} (t,
\eta_-) \hat{f} (t, \eta_+) Z(\eta_+,\eta_-) \, d\sigma= \eta\,
\widehat{Q_e^+(f,f)}(\eta) = \widehat{\left[\nabla_v
Q_e^+(f,f)\right]}(\eta)
\end{equation}
since the integral of the last term is zero. In fact, since we are
free to choose our coordinate system in the sphere, it is easy to
see that
$$
\int_{S^2} (\eta\cdot\sigma)\tilde{\eta}\, J(\eta,\sigma) \,
d\sigma = \int_{S^2} |\eta|\sigma \,J(\eta,\sigma) \, d\sigma
$$
for any function $J(\eta,\sigma)$. The desired formula
\eqref{bonn4}-\eqref{bonn5} is just the inverse Fourier transform
formula corresponding to \eqref{newfourier}. \qed

\

Now, starting from \eqref{bonn5}, we can define $H(v,w,\sigma)$ by
 \begin{align}
H(v,w,\sigma) =&\,
2\sqrt{f(w^*)}\left(\frac{3e-1}{4e} + \frac{1+e}{4e}P_{\sigma,k}\phantom{.}\right)\left[(\nabla_v \sqrt{f})(v^*)\right]\nonumber\\
&+
2\sqrt{f(v^*\phantom{.})}\left(\frac{1+e}{4e}\ - \ \frac{1+e}{4e}P_{\sigma,k}\right)\left[(\nabla_v \sqrt{f})(w^*)\right]\ ,\nonumber\\
=&\, \left[\frac{3e-1}{2e}\sqrt{f(w^*)}(\nabla_v \sqrt{f})(v^*) +
\frac{1+e}{2e} \sqrt{f(v^*\phantom{.})}
(\nabla_v \sqrt{f})(w^*)\right]\nonumber\\
&+
P_{\sigma,k}\left(\frac{1+e}{2e}\right)\left[\sqrt{f(w^*)}(\nabla_v
\sqrt{f})(v^*) - \sqrt{f(v^*\phantom{.})}
(\nabla_v \sqrt{f})(w^*)\right]\nonumber\\
=&\, H_1(v,w,\sigma) +  P_{\sigma,k}H_2(v,w,\sigma)\ ,
\label{bonn6}
\end{align}
where the last line defines $H_1(v,w,\sigma)$ and
$H_2(v,w,\sigma)$, and $G(v,w,\sigma)  = \sqrt{f(v^*)f(w^*)}
H(v,w,\sigma)$. Thus, we get
 \begin{equation}\label{bonn7}
\nabla_v Q_e^+(f,f) = \frac{1}{4\pi}\int_{S^2}\int_{\R^3} B_e^+(\sigma\cdot k) \sqrt{f(v^*)f(w^*)} H(v,w,\sigma) \dd\sigma\dd w\ .
\end{equation}
Therefore, by the Schwarz inequality
 \begin{equation}\label{bonn8}
|\nabla_v Q_e^+(f,f)(v)|  \le \left(Q_e^+(f,f)(v)\right)^{1/2} \left(
\frac{1}{4\pi}\int_{S^2}\int_{\R^3} B_e^+(\sigma\cdot k) H^2(v,w,\sigma) \dd\sigma\dd w\right)^{1/2}\ .
\end{equation}
From here we obtain a bound of $I(Q_e^+(f,f))$: Squaring both sides, and integrating in $v$ we obtain
 \begin{equation}\label{bonn9}
 I(Q_e^+(f,f)) \le
\frac{1}{4\pi}\int_{\R^3}\int_{\R^3} \int_{S^2}B_e^+(\sigma\cdot k) |H(v,w,\sigma)|^2 \dd\sigma\dd w\dd v\ .
\end{equation}
It remains to estimate the integral on the right in terms of
$I(f)$.  This consists of the sum of three terms:
\begin{align}\label{bonn21}
S_1 &= \frac{1}{4\pi}\int_{\R^3}\int_{\R^3} \int_{S^2}
B_e^+(\sigma\cdot k)| H_1(v,w,\sigma)|^2 \dd\sigma\dd w\dd v\nonumber\\
 S_2 &=
 \frac{1}{4\pi}\int_{\R^3}\int_{\R^3} \int_{S^2}
 B_e^+(\sigma\cdot k) |P_{\sigma,k}  H_2(v,w,\sigma)|^2 \dd\sigma\dd w\dd v\\
S_3 &=
 \frac{1}{4\pi}\int_{\R^3}\int_{\R^3} \int_{S^2}
 B_e^+(\sigma\cdot k) (H_1\cdot P_{\sigma,k} H_2)(v,w,\sigma) \dd\sigma\dd w\dd
 v \ .\nonumber
\end{align}
Summarizing the discussion so far, we have: For any probability
density $f$ on $\R^3$, we have
$$
I(Q_e^+(f,f)) \le S_1+S_2+2S_3\ ,
$$
where the quantities on the right hand side are specified in
(\ref{bonn21}). Our next lemma simplifies these expressions by a
change of variables:

\begin{lm}\label{ccc5}
 For any probability density $f$ on $\R^3$, we have
\begin{align}\label{bonn22}
S_1 &= \frac{1}{4\pi}\int_{\R^3}\int_{\R^3} \int_{S^2}
B(\sigma\cdot k)| F_1(v,w)|^2 \dd\sigma\dd w\dd v\nonumber\\
 S_2 &=
 \frac{1}{4\pi}\int_{\R^3}\int_{\R^3} \int_{S^2}
 B(\sigma\cdot k) |P_{\sigma',k'}F_2(v,w) |^2 \dd\sigma\dd w\dd v\\
S_3 &=
 \frac{1}{4\pi}\int_{\R^3}\int_{\R^3} \int_{S^2}
 B(\sigma\cdot k) (F_1(v,w)\cdot P_{\sigma',k'} F_2)(v,w) \dd\sigma\dd w\dd v\nonumber
 \end{align}
where
\begin{equation}\label{new2}
k'  = \frac{(1-e)k + (1+e)\sigma}{\sqrt{2(1+e^2) + 2(1-e^2)k\cdot
\sigma}} \ , \qquad \sigma'  = \frac{(1+e)k +
(1-e)\sigma}{\sqrt{2(1+e^2) + 2(1-e^2)k\cdot \sigma}}\ ,
\end{equation}
\begin{equation}\label{bonn23}
F_1(v,w) =
\left(\frac{3e-1}{2e}\sqrt{f(w)}(\nabla \sqrt{f})(v) + \frac{1+e}{2e} \sqrt{f(v)}
(\nabla \sqrt{f})(w)\right)
\end{equation}
and
\begin{equation}\label{bonn24}
F_2(v,w) =
\left(\frac{1+e}{2e}\right)\left(\sqrt{f(w)}(\nabla\sqrt{f})(v) -
\sqrt{f(v)} (\nabla \sqrt{f})(w)\right)\ .
\end{equation}
\end{lm}

\noindent{\bf Proof:} In the expressions in (\ref{bonn21}), we are
integrating over post collisional variables. We use the change of
variables Theorem \ref{backforth}, which concerns the
transformation $C_{s,e}(v,w,\sigma)\mapsto(v^*,w^*,\sigma^*)$ from
post to pre collisional variables under the ``swapping map'';
i.e., for the sigma representation. Consulting Theorem
\ref{backforth} and the definition of $H_1$ and $H_2$ in
(\ref{bonn6}), we see that each of the integrands above can be
written out in the longer form appearing in Theorem
\ref{backforth}, e.g.,
$$
| H_1(v,w,\sigma)|^2 = K[(v^*,w^*,\sigma^*),(v,w,\sigma)] =
K[C_{s,e}^{-1}(v,w,\sigma),(v,w,\sigma)] \ .
$$
Theorem \ref{backforth} allows us to write this as an integral over
$$
K[(v,w,\sigma),C_{s,e}(v,w,\sigma)] = K[(v,w,\sigma),(v',w',\sigma')] \ .
$$
Doing this for each of the three integrals in (\ref{bonn21}),
we obtain the stated formulas. \qed

\

Define the matrix $A_{\sigma,k}$ by
$$
A_{\sigma,k} = P_{\sigma',k'} - P_{k,\sigma}\ .
$$
We shall now prove:
\begin{lm}\label{ccc1} For all $\sigma$, $k$ and $z$ in $\R^3$ such that $|\sigma| = |k|=1$,
$$|A_{\sigma,k}(z)| \le 2\frac{1-e}{e}|z|\ .$$
\end{lm}

In proving this Lemma, as well as for estimating $S_2$, we shall make use of the following
lemma of Villani:

\begin{lm}\label{ced3} {\rm \cite[Lemma 4]{Vil98}}
For all $\sigma$, $k$ and $z$ in $\R^3$ such that $|\sigma| =
|k|=1$,
$$
|P_{\sigma,k}(z)| \le |z|
$$
with equality if and only if $\sigma$, $k$ and $z$ belong to the same plane.
\end{lm}

\noindent{\bf Proof of Lemma \ref{ccc1}:}, considering the
formulas for $k'$ and $\sigma'$ given in (\ref{relmom2s}) and
(\ref{relmom3s}) respectively, notice that as $e\to 1$, we have
$k' \to \sigma$ and $\sigma' \to k$, as we should, since in the
elastic case, this is what the swapping map does. Therefore, using
(\ref{relmom2s}) and (\ref{relmom3s}) we compute
$$
k' - \sigma = \frac{(1-e)k +(e+1-\sqrt{2(1+e^2) + 2(1-e^2)k\cdot\sigma})\sigma}
{\sqrt{2(1+e^2) + 2(1-e^2)k\cdot\sigma}}\ ,
$$
and
$$
\sigma' - k = \frac{(1-e)\sigma +(e+1-\sqrt{2(1+e^2) +
2(1-e^2)k\cdot\sigma})k} {\sqrt{2(1+e^2) + 2(1-e^2)k\cdot\sigma}}\
.
$$
Using the elementary estimates
$$
2e \le  \sqrt{2(1+e^2) + 2(1-e^2)k\cdot\sigma} \le 2\ ,
$$
we easily find that
$$
|k'- \sigma| \le \frac{1-e}{e}\qquad{\rm and}\qquad
|\sigma'- k| \le \frac{1-e}{e}\ .
$$
Now, notice that $P_{\sigma',k'} = P_{k,\sigma} +   P_{\sigma'
-k,k'} +  P_{k,k'-\sigma}$. This means that $A_{\sigma,k} =
P_{\sigma' - k,k'} +  P_{k,k'-\sigma}$, and now the result follows
from Lemma \ref{ced3} and the triangle inequality. \qed

\

Now we are ready to estimate $S_1$, $S_2$, and $S_3$ in terms of
$I(f)$, and prove the main result of this section.

\medskip
\noindent{\bf Proof of Theorem \ref{fgrow}:}  First of all, notice that
by Lemma \ref{ced3},
 \begin{equation}\label{bonn36}
 S_2 \le \widetilde S_2
 \end{equation}
where
 \begin{equation}\label{bonn37}
 \widetilde S_2 =
 \frac{1}{4\pi}\int_{\R^3}\int_{\R^3} \int_{S^2}
 B(\sigma\cdot k) |F_2(v,w) |^2 \dd\sigma\dd w\dd v\ .
 \end{equation}
Next, we have
\begin{multline}
S_3 =  \frac{1}{4\pi}\int_{\R^3}\int_{\R^3} \int_{S^2}
 B(\sigma\cdot k) (F_1\cdot P_{k,\sigma} F_2)(v,w,\sigma) \dd\sigma\dd w\dd v\\
 +  \frac{1}{4\pi}\int_{\R^3}\int_{\R^3} \int_{S^2}
 B(\sigma\cdot k) (F_1\cdot A_{\sigma,k} F_2)(v,w,\sigma) \dd\sigma\dd w\dd
 v\, .
 \end{multline}
Since $B$ is even in $\sigma$, $P_{k,\sigma}$ is odd in $\sigma$,
the first integral is zero. Then, by Lemma \ref{ccc1} and the
Schwarz inequality,
\begin{equation}\label{bonn30}
 S_3 \le 2\frac{1-e}{e}\sqrt{S_1\widetilde S_2} \le  \frac{1-e}{e}\left(S_1 + \widetilde S_2\right)\ .
\end{equation}
Notice that this vanishes in the elastic limit; this is the key
point discovered by Villani in the elastic case. Finally, we need
to estimate $S_1$ and $\widetilde S_2$. For this, notice that
\begin{align*}
|F_1(v,w)|^2 = \left(\frac{3e-1}{2e}\right)^2 f(w)|\nabla
\sqrt{f(v)}|^2 &+ \left(\frac{1+e}{2e}\right)^2 f(v)|\nabla
\sqrt{f(w)}|^2\\ &+  2\left(\frac{3e^2 + 2e
-1}{4e^2}\right)\sqrt{f(v)f(w)}
  \nabla\sqrt{f(v)}\cdot    \nabla\sqrt{f(v)}\ ,
\end{align*}
and
 \begin{equation*}
|F_2(v,w)|^2 =  \left(\frac{1+e}{2e}\right)^2\left[  f(w)|\nabla \sqrt{f(v)}|^2 +
  f(v)|\nabla \sqrt{f(w)}|^2 -2 \sqrt{f(v)f(w)}
  \nabla\sqrt{f(v)}\cdot    \nabla\sqrt{f(v)}\right] \ .
  \end{equation*}
Thus,
\begin{align*}
 |F_1(v,w)|^2 + |F_2(v,w)|^2 = &\, \left(\frac{5e^2-2e+1}{2e^2}\right)f(w)|\nabla \sqrt{f(v)}|^2
 + \left(\frac{e^2+2e+1}{2e^2}\right)f(v)|\nabla \sqrt{f(w)}|^2\\
 &+ \left(\frac{e^2-1}{e^2}\right)\sqrt{f(v)f(w)}
  \nabla\sqrt{f(v)}\cdot \nabla\sqrt{f(v)}\ .
\end{align*}
Now using the fact that, with our chosen normalization,
$\int_{S^2}B(k\cdot\sigma)\dd \sigma =4\pi$, together with the
Schwarz inequality and the definition of $I(f)$, we have
$$
S_1 + \widetilde S_2 \le \frac{7e^2+1}{8e^2}I(f) \ .
$$
Combining this with (\ref{bonn30}), we obtain,
$$
S_1+S_2+2S_3 \le
\left(1+2\frac{1-e}{e}\right)\frac{7e^2+1}{8e^2}I(f) = \left( 1+
(1-e)\left(\frac{2+e+15e^2}{8e^2}\right)\right)I(f)\ .
$$
This proves Theorem \ref{fgrow}. \qed

%%%%%%%%%%%%%%%%%%%%%%%%%%%%%%%%%%%%%%%%%%%%%%%%%%%%%%%%%%%%%%%%%%%%%%%%

\section{Propagation of regularity}

The next lemma relates the Fisher information bound to an
$L^\infty$-Fourier bound, similar arguments were used in
\cite{LT95,CT}. Nevertheless, we include its idea for
completeness.

\begin{lm}\label{fishbnd}
For any probability density $g$ on $\R^3$, there is a constant $C$
such that
$$
\||\eta| \hat{g}(\eta)\|_{L^\infty(\R^3)} \le C\, I(g)^{1/2}\ .
$$
\end{lm}

\noindent{\bf Proof:} Let $h = \sqrt g$. Then, the Fourier
transform of $g$ can be written as the convolution of $h$ with
itself, $ \gg(\eta) = (\hh*\hh)(\eta)$. Now, the boundedness of
the Fisher information of $g$ implies that $h=\sqrt g \in
H^1(\R^3)$, and thus
\begin{align*}
 |\eta| |\gg(\eta)| = |\eta| \left| \int_{\R^3} \hh(\eta-\eta_*)\hh(\eta_*) \, d\eta_*
 \right| &\!\le\! \int_{\R^3}\!\!\left( |\eta-\eta_*| +|\eta_*| \right)
|\hh(\eta-\eta_*)||\hh(\eta_*)|
\,d\eta_*\\[3mm]
&\!\leq \!2\left(\int_{\R^3}|\eta|^{2}|\hh(\eta)|^2 \, d\eta
\right)^{1/2}\!\!\!\left(\int_{\R^3}|\hh(\eta)|^2 \, d\eta
 \right)^{1/2}
\end{align*}
giving the desired result.\qed

\

From now on, we will restrict our attention to the most relevant
case in the literature in which $B$ is constant and due to
normalization $B=1$ or equivalently $\tilde{B}(s)=2|s|$, see
appendix and \eqref{change3}. In this case, we remind from
\eqref{Edef} that $E=(1-e^2)/8$. We shall combine Theorem
\ref{fishpro} with the following result, due to Bobylev,
Cercignani and Toscani \cite{Bobylev-Cercignani-Toscani} and Bisi,
Carrillo and Toscani \cite{BisiCT2} in this form, see also
previous results \cite{Bobylev-Cercignani2}, which gives the
uniform weak norm control. We need some notation, for any
$0<\alpha<1$, let us consider
$$
A_1(\alpha,e) = \frac{2}{4+\alpha} \left[ \left( \frac{1+e}{2}
\right)^{2+\alpha} + \frac{1- \left( \frac{1-e}{2}
\right)^{4+\alpha}}{1-\left( \frac{1-e}{2} \right)^2} \right]
$$
and $A_2(\alpha,e)=1-A_1(\alpha,e)-E\,(2+\alpha)$.

\begin{thm}\label{weaknorm} {\rm \cite[Theorem 4.6, Remark 4.7]{BisiCT2}}
For any solution $g(v,t)$ of \eqref{lis10} with constant collision
frequency $B=1$, corresponding to the initial value~$f_0$ with
unit mass, zero mean velocity such that $|v|^{2+\alpha}f_0\in L^1
(\R^3)$ with $0<\alpha<1$, there exist positive constants $C(f_0)$
and $\gamma(\alpha,e)$ such that
$$
d_2(g(t), g_\infty):=\sup_{|\eta|\neq 0}\frac{|\widehat g(\eta,t)
- \widehat g_\infty(\eta)|}{|\eta|^2} \le C \ee^{-\gamma t}
$$
for all $t\geq 0$ with
$$
\gamma(\alpha,e)=\min\left( \frac{2}{2+\alpha} A_2(\alpha,e) ,
\frac{(3-e)(1+e)}{8} \right) \ .
$$
\end{thm}

\

Let us remark that
\begin{equation}\label{new3}
\gamma(\alpha,e)\to
\gamma^*:=\min\left(\frac{2\alpha}{(2+\alpha)(4+\alpha)} , \frac12
\right) \nearrow \frac{2}{15}
\end{equation}
as $e\to 1$ and $\alpha\to 1$ respectively. Combining Theorems
\ref{fishpro} and \ref{weaknorm} and Lemma \ref{fishbnd}, we shall
prove one of our main results, Theorem \ref{mainintro}, whose
statement we now make more precise.

\begin{thm}\label{main}
For any $0<\delta<1$, there are computable positive constants $C$,
$\gamma'$, such that for any solution $g$ of \eqref{lis10}
corresponding to the initial value~$f_0$ with unit mass, zero mean
velocity, $|v|^{2+\alpha}f_0\in L^1 (\R^3)$ with $0<\alpha<1$ and
$I(f_0) < \infty$, then
$$
\sup_{\eta\in \R^3} |\eta|^{\delta} |\widehat g(\eta,t)| \le
C\ee^{-\gamma' t} + \sup_{\eta\in \R^3} |\eta|^\delta |\widehat
g_\infty(\eta,t)| = C\ee^{-\gamma' t} + C_\infty
$$
for all $t>0$, being $e$ close enough to 1.
\end{thm}

\medskip

\noindent{\bf Proof:} Pick some $R > 0$. By Lemma \ref{fishbnd},
for all $\eta$ with $|\eta| \ge R$, and all $\delta < 1$,
$$
|\eta|^\delta|\widehat g(\eta)| \le R^{\delta -1}|\eta| |\widehat
g(\eta)| \le   R^{\delta -1}C\,I(g)^{1/2}\le   R^{\delta -1}
C\ee^{c_1(e) \, t}\ ,
$$
where we used Theorem \ref{fishpro} and
$$
c_1(e)= \frac{(1-e)(2+e+15e^2)}{16e^3} - E \ .
$$
On the other hand, for $|\eta| \le R$, we have
\begin{eqnarray}
|\eta|^\delta|\widehat g(\eta)| &\le& |\eta|^\delta|\widehat g(\eta) - \widehat g_\infty(\eta)| + |\eta|^\delta  |\widehat g_\infty(\eta)|\nonumber\\
&=& |\eta|^{\delta+2}\frac{|\widehat g(\eta) - \widehat g_\infty(\eta)|}{|\eta|^2} + |\eta|^\delta  |\widehat g_\infty(\eta)|\nonumber\\
&\le& R^{\delta+2}\frac{|\widehat g(\eta) - \widehat g_\infty(\eta)|}{|\eta|^2} + |\eta|^\delta  |\widehat g_\infty(\eta)|\nonumber\\
&\le&   R^{\delta+2} C \ee^{-\gamma(\alpha,e) t} + |\eta|^\delta
|\widehat g_\infty(\eta)|\ .\nonumber
\end{eqnarray}
Combining estimates, we have that for all $\eta$,
$$
|\eta|^\delta|\widehat g(\eta)| \le R^{\delta -1}C\ee^{tc_1(e)}+
R^{\delta+2} C\ee^{-\gamma(\alpha,e) t} + |\eta|^\delta |\widehat
g_\infty(\eta)|\ .
$$
We now minimize in $R$. Up to a constant multiple, the optimal
choice is $R = \ee^{t(c_1(e) + \gamma(\alpha,e))/3}$. This results
in
$$
|\eta|^\delta|\widehat g(\eta)| \le
C\exp\left(\frac{c_1(e)(\delta+2) +(\delta - 1)
\gamma(\alpha,e)}{3} t\right)+ |\eta|^\delta |\widehat
g_\infty(\eta)|\ .
$$
Choosing $\delta < 1$ we see that for $e$ sufficiently close to
$1$, so that $c_1(e)$ is sufficiently close to $0$ and
$\gamma(\alpha,e)\simeq\gamma^*>0$, see \eqref{new3}, the exponent
is negative. Finally, taking into account the regularity obtained
by Bobylev and Cercignani for the homogeneous cooling state
$g_\infty$ in \cite[Theorem 5.3]{Bobylev-Cercignani2}, we deduce
$$
\exp\{ -|\eta|^2 \} \le \left| \hat g_\infty (|\eta|)\right| \le
\exp\{ -|\eta| \}(1+|\eta|)\ ,
$$
from which $C_\infty<\infty$. \qed

\

Now, let us proceed to write the evolution of Sobolev-type norms
for our model. Since moments in Fourier space will have simpler
relations, we shall use the homogeneous Sobolev quantities, with
$r\geq 0$, defined in \eqref{Sob_n}. Its evolution for solutions
of \eqref{lis10} is given by
\begin{align}
 \frac d{dt}\int_{\R^3}|\eta|^{2r}|\gg(\eta)|^2\, d\eta =\,& 2\int_{\R^3}\int_{S^2}
 |\eta|^{2r} \hat{g}(\eta_-) \hat{g}(\eta_+)
\hat{g}^c (\eta) \, d\sigma\, d\eta \nonumber\\*[.3cm] & -2
\int_{\R^3}|\eta|^{2r}|\gg(\eta)|^2\, d\eta - 2(2r+3)
\int_{\R^3}|\eta|^{2r}|\gg(\eta)|^2\, d\eta, \label{regstoch1}
\end{align}
where $z^c$ is the complex conjugate of $z$. Let us start by
estimating the contribution of the first term. We need to estimate
the regularity contribution of $Q_e^+(g,g)$ and for this, we make
use of the estimate of
$$
\||\eta|^\delta \hat{g}(t,\eta)\|_{L^\infty(\R^3)}
$$
obtained in Lemma \ref{fishbnd}. In fact the situation is quite
similar to \cite[Subsubsection 7.2.4]{CT} and \cite[Lemma
7.13]{CT} in the case of thermalization by a bath of particles,
adding a linear Boltzmann type operator. We will make use of the
following lemma of Carrillo and Toscani.

\begin{lm}{\rm \cite[Lemma 7.13, Proposition 7.30]{CT}}
Let $g \in \dot{H}^{r}(\R^3)$ and a probability density, then if
$$
\||\eta|^\delta \hat{g}(\eta)\|_{L^\infty(\R^3)} <\infty
$$
holds with $0<\delta <1$ and $r\geq \frac\delta2$, then
$$
\left|\int_{\R^3}\int_{S^2} |\eta|^{2r} \hat{g}(\eta_-)
\hat{g}(\eta_+) \hat{g}^c (\eta) \, d\sigma\, d\eta\right| \leq
C(r,e)\, \||\eta|^\delta \hat{g}(\eta) \|_{L^\infty(\R^3)}\, \|
g\|_{\dot{H}^{r-\delta/2}(\R^3)}^2.
$$
Here, the constant $C$ degenerates as $e\to 1$ as $C(r,e)\simeq
\left(\frac{1-e}2\right)^{-\frac{r}2-\frac14 +\frac{\delta}{4}}$.
\label{techreg}
\end{lm}

Taking into account Lemmas \ref{fishbnd}, Theorem \ref{main} and
\ref{techreg}, we deduce from the evolution of Sobolev-type norms
in \eqref{regstoch1} that
\begin{align}
 \frac d{dt} \|g\|_{\dot{H}^{r}(\R^3)}^2 \leq \, D_1\, \|g\|_{\dot{H}^{r-\delta/2}(\R^3)}^2
-4(r+2) \|g\|_{\dot{H}^{r}(\R^3)}^2 \label{regstoch2}
\end{align}
with
$$
D_1:= C(r,e)\,\sup_{t\ge 0}\||\eta|^\delta
\hat{g}(t,\eta)\|_{L^\infty(\R^3)} <\infty .
$$
We finally use standard Nash-type inequalities, see for instance
\cite[Lemma 7.14]{CT}.

\begin{lm}\label{nash}
Let $g \in \dot{H}^{r}(\R^3)$ and a probability density with
$r\geq \frac{\delta}2$, $0<\delta<1$, then $g \in
\dot{H}^{r-\delta/2}(\R^3)$ and
\begin{equation}\label{nash-n}
 \|g\|_{\dot{H}^{r}(\R^3)} \ge c_{r,\delta} \,\left( \|g\|_{\dot{H}^{r-\delta/2}(\R^3)}\right)^{(2r+3)/(2r+3-\delta)}
\end{equation}
with
\[
 c_{r,\delta} =  \left(\frac 1{2\pi}\right)^{2/(2r+3-\delta)}\left(\frac{2r+3-\delta}{2r+3}\right)^{(2r+3)/(2r+3-\delta)}.
\]
\end{lm}

\

The previous lemma allows us to obtain the inequality
\begin{align}
 \frac d{dt} \|g\|_{\dot{H}^{r}(\R^3)}^2 \leq \, D_2\,
 \left[\|g\|_{\dot{H}^{r}(\R^3)}^2\right]^{\theta}
-4(r+2) \|g\|_{\dot{H}^{r}(\R^3)}^2 \label{regstoch3}
\end{align}
with $D_2$ easily obtained from above and
$\theta=(2r+3-\delta)/(2r+3)<1$. As a consequence, we achieve one
of the main theorems of our work.

\begin{thm}\label{propreg}
Given the solution $g$ of \eqref{lis10} corresponding to the
initial value~$f_0\in \dot{H}^{r}(\R^3)$, with $r>0$, of unit
mass, zero mean velocity such that $|v|^{2+\alpha}f_0\in L^1
(\R^3)$ with $0<\alpha<1$ and $I(f_0) < \infty$. Then, for $e$
close to 1, the solution~$g(t,v)\!$ of~\eqref{lis10} is bounded
in~$\dot{H}^r(\R^3)$, and there is a universal constant~$A$ so
that, for all $t>0$,
\[
\|g(t) \|_{\dot{H}^r(\R^3)} \le  \max\left\{\|f_0
\|_{\dot{H}^r(\R^3)}, A\right\}.
\]
In particular, the stationary solution or homogenous cooling
profile $g_\infty$ to \eqref{lis10} belongs to $H^\infty(\R^3)$.
\end{thm}

\begin{remark}
Let $f_0\in\dot{H}^r(\R^3)$, with $r>0$ of unit mass, zero mean
velocity such that $|v|^{2+\alpha}f_0\in L^1 (\R^3)$ with
$0<\alpha<1$ and $I(f_0) < \infty$, be any initial datum for
equation~\eqref{lis10}. Previous theorem together with the Nash
inequality in Lemma {\rm\ref{nash}} implies that, for $e$ close to
1, the solution~$g(t,v)\!$ of~\eqref{lis10} is bounded
in~$L^2(\R^3)$, and there is a universal constant~$C_2$ so that,
for all $t>0$,
\[
\|g(t) \|_{L^2(\R^3)} \le  \max\left\{\|f_0
\|_{\dot{H}^r(\R^3)}^{3/(3+2r)}, C_2\right\}.
\]
\end{remark}

Let us point out that the previous propagation of smoothness
results are true for any value of $e$ for which a uniform in time
estimate of $ \||\eta|^\delta \hat{g}(t,\eta)\|_{L^\infty(\R^3)}$
is available, which in our case is given by the values of $e$ for
which the estimate in Theorem \ref{fishpro} is satisfied with
$\gamma'\ge 0$.

Finally, using the strategy already introduced in \cite{CGT} and
used in inelastic models in \cite{BisiCT,CT}, see also
\cite{Vil06}, we can obtain the convergence in $L^1$. The first
ingredient is an interpolation inequality that allows to control
distances in arbitrary Sobolev norms and in $L^2$ by using the
propagation of smoothness and the convergence result in Theorem
\ref{weaknorm} in \cite[Theorem 4.6, Remark 4.7]{BisiCT2}.

\begin{prop}\label{d2tosobolev}
{\rm \cite[Theorem 4.1]{CGT}} Let $s\ge 0$, and $\beta_1
>0$, $0<\beta_2<1$ be given. Then
\[
\|f-g\|_{\dot{H}^s(\R^3)} \le C(\beta_1,\beta_2)\,
d_2(f,g)^{(1-\beta_2)} \min\left(\|f-g\|_{\dot{H}^{r_1}(\R^3)} ,
\|f-g\|_{\dot{H}^{r_2}(\R^3)}\right)^{\beta_2},
\]
with
\begin{gather*}
r_1 = \frac{s +2(1-\beta_2)}{\beta_2} \qquad , \qquad r_2 =
\frac{2s +(7+\beta_1)(1-\beta_2)}{2\beta_2},\\
C(\beta_1,\beta_2) = \left(
\frac{4\pi}{3}(1+3/\beta_1)\right)^{1-\beta_2},
\end{gather*}
and
$$
d_2(f,g) = \sup_{|\eta| \neq 0}
\frac{|\hat{f}(\eta)-\hat{g}(\eta)|}{|\eta|^2}\ .
$$
\end{prop}

The previous result implies immediately convergence in strong
norms:

\begin{thm}\label{hrconv}
Let $g$ be the solution of \eqref{lis10} corresponding to the
initial probability distribution function~$f_0\in
\dot{H}^{r+\epsilon}(\R^3)$, with $r>0$ and $\epsilon>0$, of zero
mean velocity such that $I(f_0) < \infty$. Then, for $e$ close to
1, the solution~$g(t,v)\!$ of~\eqref{lis10} converges strongly in
$\dot{H}^r$ with an exponential rate towards the homogenous
cooling state, i.e., there exist positive constants $C$ and
$\tilde\gamma$ explicitly computable such that
$$
\|g(t)-g_\infty \|_{\dot{H}^r(\R^3)} \leq C\, \ee^{-\tilde\gamma
t}
$$
for all $t>0$.
\end{thm}

Let us point out that the exponential rate $\tilde \gamma$ can be
computed as $\tilde\gamma=(1-\beta_2)\gamma$ for any choice of
$\beta_1>0$, $0<\beta_2<1$ such that $\max(r_1,r_2)<r+\epsilon$.
Moreover, the uniform control of moments for the probability
measure yields control of the distance in $L^1$.

\begin{lm}\label{L2mtoL1}
{\rm \cite[Theorem 4.2]{CGT}} Let $f\in L^1\cap L^2(\R^3)$ with
$|v|^{2p}f\in L^1(\R^3)$, then, for all $p>0$,
\[
\int_{\R^3} |f(v)|\, dv   \le C(p)\left(  \int_{\R^3} |f(v)|^2 \,
dv \right)^{2p/(3+4p)} \left(  \int_{\R^3} |v|^{2p}|f(v)| \, dv
\right)^{3/(3+4p)}
\]
with
\[
C(p) = \left[ \left(\frac 3{4p} \right)^{4p/(3+4p)} + \left(\frac
{4p}3
\right)^{3/(3+4p)}\right]\left(\frac{4\pi}{3}\right)^{2p/(3+4p)}.
\]
\end{lm}

Let us recall what is known about tails of the homogeneous cooling
state. One very interesting property is that not all moments of
$g_\infty$ are bounded and the threshold moment depends on the
restitution coefficient $e$. This was proved by
\cite{Bobylev-Cercignani2}, see
\cite{Bobylev-Cercignani-Gamba,Bobylev-Cercignani-Gamba2} for
generalizations. In particular, the fourth moment of $g_\infty$ is
bounded for all restitution coefficients.

The next ingredient we need to pass from $L^2$ to $L^1$
convergence is the uniform-in-time propagation of the fourth
moment for solutions for any value of the restitution coefficient
as obtained in \cite[Appendix]{Bolley-Carrillo} from which the
main Theorem \ref{l1conv} immediately follows.

The exact value of the constant $\gamma'$ depends on the value of
$r$ in the hypotheses of Theorem \ref{l1conv}. In fact, by taking
$s=0$ in Proposition \ref{d2tosobolev}, we get that the decay in
$L^2$ will be given by a constant $\tilde\gamma=\gamma(1-\beta_2)$
for any choice of $\beta_1>0$, $0<\beta_2<1$ such that
$$
\max\left( \frac{2(1-\beta_2)}{\beta_2} ,
\frac{(7+\beta_1)(1-\beta_2)}{2\beta_2}\right)<r.
$$
Once, we have this decay of the $L^2$ norm, the previous lemma
finally gives the value $\gamma'=\frac{8}{11}\tilde\gamma$.

%%%%%%%%%%%%%%%%%%%%%%%%%%%%%%%%%%%%%%%%%%%%%%%%%%%%%%%%%%%%%%%%%%%%%%%%

\section{Small inelasticity limit of HCS}

As a further application of the results proven in the previous
sections, we study the small inelasticity limit of the homogeneous
cooling states and prove, as one might expect, that as $e\to 1$
then the homogeneous cooling state converges towards the
corresponding Maxwellian in strong norms. Previous results in this
direction were done in the asymptotic expansion in Fourier for the
self-similar solution, see \cite[Subsection
6.1]{Bobylev-Carrillo-Gamba}.

Let us fix for any small $\varepsilon=\frac{1-e}2$, the
corresponding unique smooth $g_\infty^\varepsilon\in
H^\infty(\R^3)$ stationary state to \eqref{lis10} with zero mean
velocity and temperature fixed by the initial data. Then, we can
show the following result:

\begin{thm}\label{smallinelasticity}
Given $M$ the Maxwellian with zero mean velocity and temperature
given by the initial temperature of $f_0$, then there exist a
positive constant $C$ such that
$$
\|g_\infty^\varepsilon-M\|_{L^1(\R^3)} \leq C \varepsilon^{1/2}
\left[1 + |\log \varepsilon|^{1/2}\right] ,
$$
for any $\varepsilon>0$ small enough.
\end{thm}

\medskip

\noindent{\bf Proof:} Let $g^\varepsilon(t)$ be the solution to
\eqref{lis10} with initial data $M$, then
$$
\|g_\infty^\varepsilon-M\|_{L^1(\R^3)} \leq
\|g^\varepsilon(t)-g_\infty^\varepsilon\|_{L^1(\R^3)} +
\|g^\varepsilon(t)-M\|_{L^1(\R^3)}.
$$
Now, we are going to control each term separately. Since $M\in
H^\infty(\R^3)$, then $g^\varepsilon(t)$ will satisfy due to
Theorem \ref{l1conv} that
$$
\|g^\varepsilon(t)-g_\infty^\varepsilon \|_{L^1(\R^3)} \leq
C\,\varepsilon^{-r}\, \ee^{-\gamma' t}
$$
for all $t>0$. Here, we have made explicit the dependence on the
restitution coefficient of the constants in the previous section.
Actually, revising the discussion on the value of the constants in
the previous section, one gets that $\gamma'$ can be made as close
to $\frac{8}{11}\gamma$ as we want since our solution
$g^\varepsilon(t)$ lies in $H^\infty(\R^3)$ due to Theorem
\ref{propreg}, see last paragraph of the previous section.
Moreover, we can fix $\varepsilon$ small enough and $\alpha$ close
enough to 1 in such a way that $\gamma$ is as close as we want to
$2/15$ in \eqref{new3}. For example, by choosing $1/11$, then for
$\alpha\simeq 1$ and for small enough $\varepsilon$ we have $1/11
< \gamma'< 16/165= 2/15 \cdot 8/11$, where $1/11$ is an arbitrary
number.

Concerning the behavior of the constants in front of the
exponential in time function as $\varepsilon\to 0$, it is not
difficult, but tedious, to check that it does not degenerate to 0
and is uniformly bounded as $\varepsilon\to 0$ in the case of the
constants for the $d_2$ distance in Theorem \ref{weaknorm} and in
Theorem \ref{main}. However, the dependence on the restitution
coefficient of the estimates in \cite[Lemma 7.13, Proposition
7.30]{CT} leading to Lemma \ref{techreg} degenerates as
$\varepsilon\to 0$ as $\varepsilon^{-r}$ with an exponent $r$
related to the regularity needed in the interpolation Proposition
\ref{d2tosobolev}. One can estimate this degeneracy exponent $r$
exactly depending on the regularity needed for having $1/11 <
\gamma'$, but it is not important its exact value as we shall see
below.

On the other hand, we can use the Csiszar-Kullback inequality
\cite{Csi67,Kul51} together with the Logarithmic Sobolev
inequality \cite{Gro75,Tos} to get:
$$
I(g^\varepsilon(t))-I(M)\geq \int_{\R^3}
g^\varepsilon(t,v)\ln\frac{g^\varepsilon(t,v)}{M(v)} \,dv \ge
\frac12 \|g^\varepsilon(t)-M \|_{L^1(\R^3)}^2\,.
$$
Using Theorem \ref{fishpro}, we deduce
$$
\|g^\varepsilon(t)-M \|_{L^1(\R^3)} \leq \left[
2\left(\ee^{\varepsilon\, \omega t}-1\right)I(M) \right]^{1/2}
$$
with
$$
\omega=\frac{2+e+15e^2}{4e^3} - \frac{1+e}2 \simeq \frac72
$$
as $e\to 1$.

Finally, for suitable choice of $\alpha$ and for $\varepsilon$
small enough, we conclude
$$
\|g_\infty^\varepsilon-M\|_{L^1(\R^3)} \leq \left[
2\left(\ee^{\frac72 \varepsilon\, t}-1\right)I(M) \right]^{1/2} +
C\,\varepsilon^{-r}\, \ee^{-\frac{1}{11} t}
$$
for all $t\geq 0$, and thus by Taylor's theorem, we get
$$
\|g_\infty^\varepsilon-M\|_{L^1(\R^3)} \leq \left[ 7 \varepsilon\,
t \, \ee^{\frac72 \varepsilon\, t} \,I(M) \right]^{1/2} +
C\,\varepsilon^{-r}\, \ee^{-\frac{1}{11} t}
$$
for all $t\geq 0$. By choosing $t=11(1/2+r)|\log \varepsilon|$, we
obtain
$$
\|g_\infty^\varepsilon-M\|_{L^1(\R^3)} \leq \left[ 77(1/2+r)
\varepsilon\, |\log \varepsilon| \, \ee^{\frac{77}{2}(1/2+r)
\varepsilon\, |\log \varepsilon|} \,I(M) \right]^{1/2} + C\,
\varepsilon^{1/2}
$$
from which the announced result follows.\qed

%%%%%%%%%%%%%%%%%%%%%%%%%%%%%%%%%%%%%%%%%%%%%%%%%%%%%%%%%%%%%%%%%%%%%%%%%%%%%%

\section*{Appendix: The kinematics of inelastic collisions}\label{kinematics}
\setcounter{equation}{0} \setcounter{thm}{0}
\def\theequation{{A.\arabic{equation}}}
\def\thethm{{A.\arabic{thm}}}
\def\thesubsection{{A.\arabic{subsection}}}

Here, we will review in detail the collision mechanism for
inelastic collisions and the weak and strong formulation in two
useful representations of the inelastic gain collision operator.
We will perform in detail the relations between the collision
frequencies in the different representations and for general
interactions being of Maxwell type or not. Basically, these
results make a summary of already known relations in particular
cases written in \cite{Bobylev-Carrillo-Gamba,GPV} but we believe
this summary sets up the more general case once and for all.

\subsection{The kinematics of elastic collisions}\label{elaskinematics}

We begin by reviewing two ways of parameterizing the set of all
elastic collisions in $\R^3$. The presentation has some unusual
features that will be useful to our investigation of inelastic
collisions. If particles with like masses and with velocities $v$
and $w$ collide elastically, so that both  energy and momentum are
conserved, then
$$
\frac{v+w}{2}\qquad{\rm and}\qquad |v-w|
$$
are both conserved. The conservation of the first quantity
directly expresses the conservation of momentum (since the masses
are the same), and then the conservation of the second follows
from the conservation of energy and the parallelogram law:
$$
\frac{|v|^2 + |w|^2}{2}  = \left|\frac{v+w}{2}\right|^2 +  \left|\frac{v-w}{2}\right|^2\ .
$$
Therefore, let us introduce
\begin{equation}\label{zukdefs}
z = \frac{v+w}{2} \qquad u = v -w\qquad{\rm and}\qquad k =
\frac{u}{|u|}\ .
\end{equation}
Note that
\begin{equation}\label{zukrec}
v = z+ \frac{|u|}{2}k \qquad{\rm and}\qquad w =  z-
\frac{|u|}{2}k\ .
\end{equation}

Now consider an elastic collision $(v,w) \longrightarrow (v',w')$
where $v'$ and $w'$ are the post collisions velocities of the two
particles. We define $z'$, $u'$ and $k'$ in terms of the post
collisions velocities $v'$ and $w'$ just as  we defined $z$, $u$
and $k$ in terms of $v$ and $w$ in (\ref{zukdefs}).  By the
conservation of $z$ and $|u|$ and (\ref{zukrec}), we have that
\begin{equation}\label{zukrec2}
v' = z+ \frac{|u|}{2}k' \qquad{\rm and}\qquad
w'=z-\frac{|u|}{2}k'\ .
\end{equation}
That is, by the conservation laws, only $k$ changes, and the
outcome of the collision is entirely specified by giving the
change in  the unit vector $k \longrightarrow k'$, together with
the initial velocities $v$ and $w$. Hence the space of
kinematically possible collisions is the set $\Xi = \R^3\times
\R^3\times S^2$ with generic point $(v,w,\omega)$. The vector
$\omega$ is called the  {\it collision vector}, and it is the
additional parameter, beyond $v$ and $w$, needed to specify the
post collisional velocities  $v'$ and $w'$. This specification is
then described by a bijective map $C$ from $\Xi$ onto itself with
$C(v,w,\omega) = (v',w',\omega')$, where $v'$ and $w'$ are the
post collisional velocities of the two particles, and $\omega$ and
$\omega'$ are  the  collision vectors that are needed to determine
$k'$ from $v$ and $w$, or in reverse, to determine $k$ from $v'$
and $w'$.

The map $C$ is called the {\em collision map}. There are two
collision maps that are particularly useful for our purposes: the
{\em swapping map} and the {\em reflection map}. The first has
many mathematical advantages, due to its simplicity, but the
latter has a closer connection with the physics of the collision
process, and we shall need them both.

\subsubsection{Reflection map:} 
\begin{figure}[ht]
\centering
\includegraphics[width=4in]{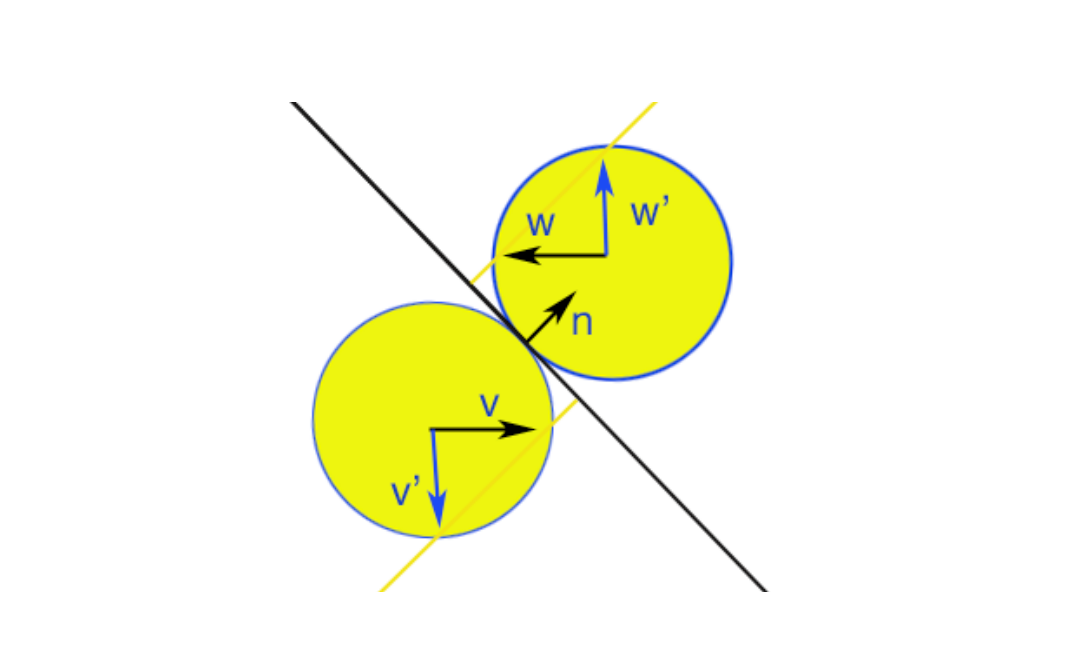}
\caption{Profile of the plane of reflection, with some normal
lines indicated, along which the pre and post collisional
velocities line up.}
\label{colelas}
\end{figure}

Consider a collision of two
identical hard spheres in the center of momentum frame, so that $z
=0$. Let $n$ be the unit vector pointing from the center of one
particle to the center of the other at the moment of collision.
(It does not matter from which to which; our expressions will be
quadratic in $n$.)  The result of the collision is exactly as if
each particle undergoes specular reflection upon striking the
plane through the point of contact that has unit normal $n$. As a
result, the velocities of both particles are reflected about this
plane, as shown in Figure \ref{colelas}.
Clearly, the relative velocity $u= v-w$ of the two particles is
also reflected about this plane, so that we have $u' = u -
2(u\cdot n)n$. This remains true if we translate out of the center
of momentum frame.  Also,  since reflection is an isometry, $k' =
k - 2(k\cdot n)n$. We therefore let $n$ stand in the role of
$\omega$, and we have
\begin{equation}\label{refdef}
k' =k - 2(k\cdot n)n \qquad{\rm and}\qquad n' = n\ .
\end{equation}
From this and (\ref{zukrec2}) we obtain the full collision map:
\begin{equation}\label{reflect}
\begin{cases} v'  &=\  v -((v-w)\cdot n) n\\[1mm]
w'  &=\   w + ((v-w)\cdot n) n\\[1mm]
n'  &=\    n
\end{cases}
\end{equation}
Let us use $C_r$ to denote the collision map $C_r(v,w,n) =
(v',w',n')$ based on (\ref{reflect}).  Of course, $n$ and $-n$
lead to the same reflection, so that if we hold $v$ and $w$ fixed
and vary $n$, we get a double cover of the set of possible post
collisional velocities $(v',w')$ resulting from $(v,w)$. However,
as is clear from the third line, as a transformation from $\Xi$
onto itself, $C_r$ is a bijection.

The next thing to observe is that $C_r$ is a measure preserving
transformation on $\Xi$. Indeed, since the measure $\dd v\dd w\dd
n$ on $\Xi$ is also the measure   $\dd z\dd u\dd n$, going to
polar coordinates in $u$, we have $\dd v\ \dd w\ \dd n =  \dd z\
\dd |u|\ \dd k \ \dd n$. Here $\dd n$ and $\dd k$ are both  the
uniform measures on $S^2$ with total mass $4\pi$, which is
consistent with our use of polar coordinates. Since it is clear
from Fubini's Theorem that $\dd k \dd n = \dd k'\dd n'$, we have
that $\dd v'\dd w'\dd n'  = \dd v\dd w\dd n$. This gives us the
collision kernel in the reflection representation
\begin{equation}\label{colkerref}
\langle Q(f,f), \varphi\rangle = \frac{1}{4\pi}
\int_{\R^3}\int_{\R^3}\int_{S^2} f(v) f(w)[\varphi(v')-
\varphi(v)] \widetilde\Phi(|u|) \widetilde B(k\cdot n) \dd n \dd v
\dd w \ .
\end{equation}
Here, $\widetilde\Phi$ and $\widetilde B$ takes into account the
difference rates of collisions and its dependence with respect to
the strength of the relative velocity and the collision angle
respectively and $\varphi$ is any test function.

Finally, since reflections are their own inverses, one sees that
$C_r$ is its own inverse. Therefore, we may regard $(v',w')$ as
pre collisional velocities that may result in the pair $(v,w)$ of
post collisional velocities. Thus, in the ``gain term''  part of
the integral for $Q(f,f)$, one can change variables
$$
\int_{\R^3}\!\!\int_{\R^3}\!\!\int_{S^2} f(v) f(w)\varphi(v')
\widetilde B(k\cdot n) \dd n \dd v \dd w =
\int_{\R^3}\!\!\int_{\R^3}\!\!\int_{S^2} f(v') f(w')\varphi(v)
\widetilde\Phi(|u'|) \widetilde B(k'\cdot n') \dd n' \dd v' \dd
w'\ .
$$
Now, recall that $\widetilde B$ is even, and  note that $k'\cdot
n' = - k\cdot n$ and $|u'|=|u|$. Using this and the measure
preserving property, we obtain
$$
\int_{\R^3}\!\!\int_{\R^3}\!\!\int_{S^2} f(v') f(w')\varphi(v)
\widetilde\Phi(|u'|) \widetilde B(k'\cdot n') \dd n' \dd v' \dd w'
= \int_{\R^3}\!\!\int_{\R^3}\!\!\int_{S^2} f(v') f(w')\varphi(v)
\widetilde\Phi(|u|) \widetilde B(k\cdot n) \dd n \dd v  \dd w\ .
$$
Thus, we have
$$
\langle Q(f,f), \varphi\rangle = \frac{1}{4\pi}
\int_{\R^3}\int_{\R^3}\int_{S^2}[f(v')f(w') - f(v) f(w)]\varphi(v)
\widetilde\Phi(|u|) \widetilde B(k\cdot n) \dd n \dd v \dd w \ ,$$
which allows us to write the collision kernel in the strong form:
\begin{equation}\label{colkerrefst}
Q(f,f)(v) = \frac{1}{4\pi} \int_{\R^3}\int_{S^2}[f(v')f(w') - f(v)
f(w)]\widetilde\Phi(|u|)\widetilde B(k\cdot n) \dd n  \dd w\ .
\end{equation}
Note that to specify everything needed to compute $Q(f,f)$, one
needs only the first two lines of (\ref{reflect}), and this is all
that is usually written down in discussions of elastic collisions.
However, in the inelastic case, the collisions are not reversible,
since they dissipate energy, and so there is not such a simple
relation between the pre and post collisional velocities. Thus,
more care is required in the passage from the weak form of the
collision kernel to the strong form, and it will be helpful to
keep all three lines in (\ref{reflect}), and remember that $C_s$
is a measure preserving bijection of $\Xi$ onto itself. Indeed,
note that for fixed $n$, the map $(v,w) \mapsto (v',w')$ is a two
to one, so dropping the third line, we would not have a bijection.

\subsubsection{Swapping map:}
As we have mentioned, there is another collision map which leads
to a different way of writing down the collision kernel. This
other way is less directly connected with the physics of hard
sphere collisions, but it does have considerable mathematical
advantages due to its simplicity, which is based on a simple
swapping of $\omega$ and $k$. In this context, it is traditional
to write $\sigma$ in place of $\omega$, and the very simple rule
for computing $k'$ and $\sigma'$ in terms of  $v$, $w$ and
$\sigma$ is simply
\begin{equation}\label{swapdef}
k' = \sigma \qquad{\rm and}\qquad \sigma' = k\ .
\end{equation}
As with (\ref{refdef}), we use this and (\ref{zukrec2}) to obtain
the corresponding collision map:
\begin{equation}\label{sigma}
\begin{cases} v'  &= {\displaystyle \frac{v+w}{2} +
\frac{|v-w|}{2}\sigma}\\[2mm]
w'  &= {\displaystyle \frac{v+w}{2} - \frac{|v-w|}{2}\sigma}\\[2mm]
\sigma'  &=  k= {\displaystyle \frac{v-w}{|v-w|}}
\end{cases}
\end{equation}
Let us use $C_s$ to denote the swapping map $C_s(v,w,\sigma) =
(v',w',\sigma')$. Like $C_r$, $C_s$ is a measure preserving
transformation on $\Xi$. As before, we note that the measure $\dd
v\dd w\dd \sigma$ on $\Xi$ is also the measure $\dd z\dd u\dd
\sigma$, and  going to polar coordinates in $u$, we have $\dd v\
\dd w\ \dd \sigma =  \dd z\ \dd |u|\ \dd k \ \dd \sigma$. In this
form it is clear that the collision map for swapping is a measure
preserving map, $\dd v'\dd w'\dd \sigma'  \ =\  \dd v\dd w\dd
\sigma$. This gives us the collision kernel in the swapping
representation:
\begin{equation}\label{wgs}
\langle Q(f,f), \varphi\rangle = \frac{1}{4\pi}
\int_{\R^3}\int_{\R^3}\int_{S^2} f(v) f(w)[\varphi(v') -
\varphi(v)] \Phi(|u|) B(k\cdot \sigma) \dd\sigma \dd v \dd w
\end{equation}
where $\varphi$ is any test function, and as above $k$ denotes the
unit vector in the direction of $v-w$, and $\Phi$ and $B$ gives
the relative rates of the various kinematically possible
collisions with respect to the strength of the relative velocity
and the collision angle, and $\dd \sigma$ is the uniform measure
on $S^2$ with total mass $4\pi$.

Note that like $C_r$, $C_s$ is its own inverse, simply because the
map in  (\ref{swapdef}) is its own inverse. Thus, we may once more
regard $(v',w')$ as a pair of pre collisional velocities. Changing
variables, and using $k'\cdot \sigma'= k\cdot \sigma$, and the
measure preserving property, we obtain the strong form
\begin{equation}\label{colkerswapst}
Q(f,f)(v) = \frac{1}{4\pi} \int_{\R^3}\int_{S^2}[f(v')f(w') - f(v)
f(w)] B(k \cdot \sigma) \dd \sigma  \dd w\ .
\end{equation}

Note that only the first two lines of  (\ref{sigma}) are required
to compute $Q(f,f)$, and this is all that is usually written down
in discussions of elastic collisions. However, it is worth noting
explicitly that if one holds $\sigma$ fixed, and then just
considers the transformation $(v,w) \mapsto (v',w')$ described by
the first two lines, this transformation is {\em not} onto and
{\em not} injective: While the direction of $u$ is arbitrary, the
direction of $u'$ is always that of $\sigma$.  In the inelastic
case, everything will be clear if we always keep in mind that the
collision map is a bijection from $\Xi$ onto itself, and keep all
three lines needed to specify this map.

There is one more important point to notice: The integral in
(\ref{wgs}) is unchanged if we swap $v$ and $w$; i.e., make the
change of variables $(v,w,\sigma) \mapsto (w,v,\sigma)$. This
transformation does not affect $v'$ but it reverses the sign on
$k$, and so we may replace $B(k\cdot\sigma)$ in  (\ref{wgs}) with
$B(-k\cdot\sigma)$ without affecting  $\langle Q^+(f,f),
\varphi\rangle$. That is, only the symmetric part of $B$
contributes to the collision kernel, and we may freely require, by
symmetrizing if need be, that $B$ is a symmetric function on
$[-1,1]$. {\em We shall always impose this symmetry requirement on
our rate functions $B$.}

\subsubsection{Relations between representations}

Our final business in this section is to relate these two
representation of the gain term, and to determine in particular
the relation between $\Phi$, $B$ and $\tilde \Phi$, $\tilde B$ in
(\ref{wgs}) and (\ref{colkerref}). The crucial fact is that both
maps $(v,w,n)\mapsto (v',w',n')$ and  $(v,w,\sigma)\mapsto
(v',w',\sigma')$ yield the same pair $(v',w')$ if $\sigma$ and $n$
are related through $\sigma = k - 2(k\cdot n)n$; i.e., if $\sigma$
is the reflection of $k$ about the plane normal to $n$. Indeed,
since, for example,
 \begin{equation}\label{bonn1}
  \frac{v+w}{2} +  \frac{v-w}{2} - ((v-w)\cdot n) n\  =\  v - ((v-w)\cdot n) n\ ,
  \end{equation} the first lines of (\ref{sigma}) and (\ref{reflect}) coincide
if we relate $n$ and $\sigma$ through
\begin{equation}\label{rel}
\frac{u}{2} - (u\cdot n)n = \frac{|u|}{2}\sigma\qquad{\rm or,\
equivalently}\qquad k - 2(k\cdot n)n = \sigma\ .
\end{equation}
Since $\sigma$ is the reflection of $k$ about the plane orthogonal
to $n$, we recover $n$ (up to a sign) in terms of $\sigma$ and $k$
through:
\begin{equation}\label{rel2}
n= \frac{u - |u|\sigma}{|u - |u|\sigma|} \qquad{\rm or,\
equivalently}\qquad n= \frac{k - \sigma}{|k - \sigma|}\ .
\end{equation}
The fact that we only recover $n$ up to a sign does not matter; it
is not $n$ itself, but only the plane normal to $n$ that matters
in all computations we shall make. Finally, from (\ref{rel}) doing
the $k\cdot$ operation over the formula for $k$, we have
\begin{equation}\label{rel3}
|k\cdot n| = \sqrt{\frac{1-k\cdot\sigma}{2}}\ .
\end{equation}
These formulas will be very useful in relating the $n$
representation and the $\sigma$ representation for inelastic
collisions.

To work out the relation between $\Phi$, $B$ and $\tilde \Phi$,
$\widetilde B$, we need just one more identity
\cite{Boby75,Bobylev}: For any test functions $\varphi$ on $\R^3$,
\begin{equation}\label{change}
\int_{S^2}\varphi\left(\frac{u - |u|\sigma}{2}\right)\dd \sigma =
\int_{S^2}\frac{2|u\cdot n|}{|u|}\varphi\left((u\cdot
n)n\right)\dd n\ .
\end{equation}
To see this, observe that if we define $\theta = \cos^{-1}(k\cdot
\sigma)$ and $\chi =  \cos^{-1}(k\cdot n)$, then from the
reflection relation between $k$, $n$ and $\sigma$ in (\ref{rel}),
we see that for $0 \le \chi \le \pi/2$, $\theta =\pi - 2\chi$, so
that
$$
\sin(\theta)\dd \theta = 2\sin(2\chi)\dd \chi = 4\cos(\chi)\sin(\chi)\dd \chi = 4|k\cdot n| \sin(\chi)\dd \chi\ .
$$
Thus, using (\ref{rel3}),
\begin{equation}\label{rel4}
\dd \sigma = 4|k\cdot n|\dd n \qquad{\rm and}\qquad \dd n =
\frac{1}{4}\sqrt{\frac{2}{1-k\cdot \sigma}}\dd \sigma\ .
\end{equation}
Finally, by taking into account that with $k$ fixed, $n\mapsto
\sigma$ is a two to one cover of $S^2$, and since $(u -
|u|\sigma)/2 = (u\cdot n)n$ as noted in (\ref{rel}), we obtain the
identity (\ref{change}).

Now since $v'$ can be expressed either in terms of $(u -
|u|)\sigma/2$ or $(u\cdot n)n$, by equating the right hand sides
of (\ref{wgs}) and (\ref{colkerref}) which requires
$\Phi=\tilde\Phi$ and $B(k\cdot\sigma)\dd \sigma = 2\widetilde
B(k\cdot n)\dd n$ since $n\mapsto \sigma$ is a two to one cover of
$S^2$ and $\tilde{B}$ is symmetric, we obtain from (\ref{rel4})
that
 \begin{equation}\label{change2}
 B(k\cdot\sigma) =\frac{ \widetilde B\left(k\cdot n\right)}{\phantom{.}2|k\cdot n|}\ .
 \end{equation}
Then since from (\ref{rel}) we have $k\cdot\sigma = 1 - 2(k\cdot
n)^2$, we obtain
\begin{equation}\label{change3}
 B(s) = \frac{
 \widetilde B\left(\sqrt{(1-s)/2}\right)}{2\sqrt{(1-s)/2}} \qquad{\rm and}\qquad
 \widetilde B(t) =  2|t| B(1-2t^2)\ ,
\end{equation}
where  $-1\le s \le 1$ and $-1 \le t \le 1$.

\subsection{The kinematics of inelastic collisions}\label{kinin}

The kinematics of inelastic collisions is easiest to describe
starting from the $n$ representation. As in the elastic case,
viewed in the center of momentum frame, the collision behaves like
a collision with a wall running normal to the direction vector $n$
pointing from the center of one ball to the center of the other
ball at the point of contact. However, in the case of an inelastic
collision with a wall, the component of the reflected velocity
normal to the wall is reduced, while the component parallel to the
wall is unchanged.  Thus the rule for updating the relative
velocity after the collision is this: Let $P_n$ denote the
orthogonal projection onto the span of $n$. Then
$$
P_n(v'-w') = -eP_n(v-w) \qquad \mbox{and} \qquad P_n^\perp(v'-w')
= P_n^\perp (v-w)\ .
$$
Thus, our rule for updating the relative velocity after the
collision is a sort of reduced reflection
$$
v'-w' = (v-w) -  (1+e)((v-w)\cdot n)n\ ,
$$
where the {\it restitution coefficient} $e$ satisfies $0 \le e \le
1$. If $e=1$, then this is a reflection about the plane normal to
$n$. If $e=0$, this simply cancels out the component of $v-w$
along $n$, which corresponds to a perfectly inelastic collision.
\begin{figure}[ht]
\centering
\includegraphics[width=4in]{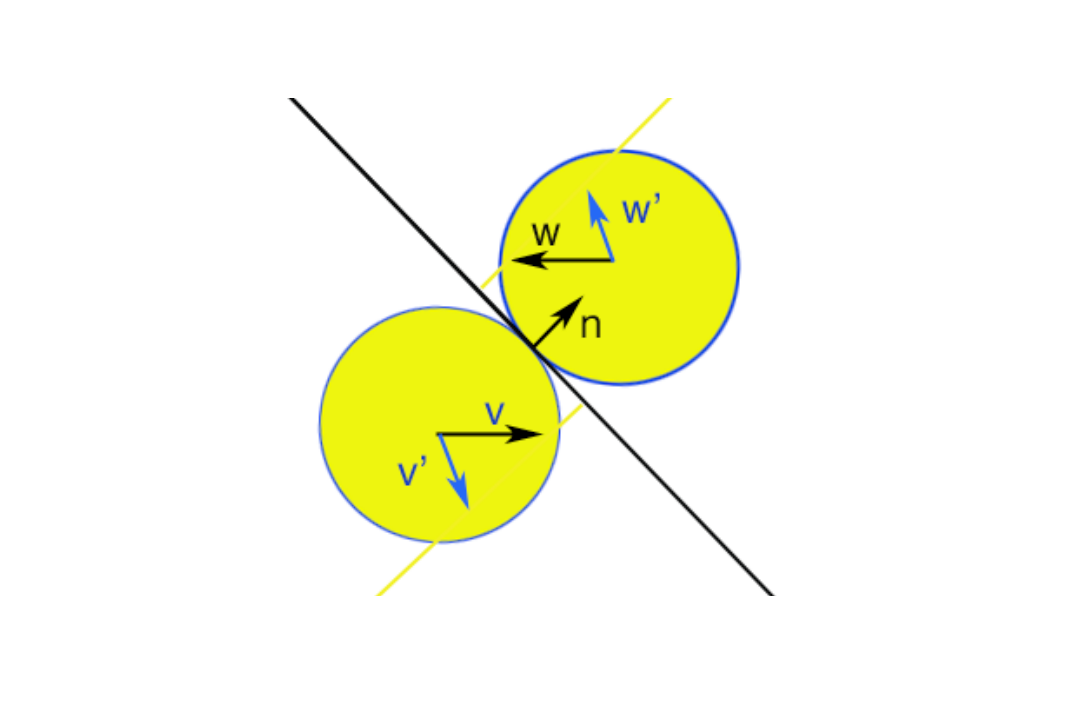}
\caption{Profile of the plane of reflection, with some normal
lines indicated, along which the pre and post collisional
velocities line up in the inelastic case.}
\label{colinelas}
\end{figure}
This gives us the {\em inelastic reflection map  with restitution
coefficient} $e$, denoted $C_{r,e}$: $C_{r,e}(v,w,n) = (v',w',n')$
where
 \begin{equation}\label{inreflect}
\begin{cases}  v'  &=\  v - {\displaystyle\frac{1+e}{2}((v-w)\cdot n)
n}\\[1mm]
 w'  &=\   w +  {\displaystyle \frac{1+e}{2}((v-w)\cdot n) n}\\[1mm]
 n'  &=\   n
\end{cases}
\end{equation}
Note that with $u' = v'-w'$, we have
\begin{equation}\label{energy1}
|u'|^2 = |u|^2 + (e^2-1)(u\cdot n)^2 .
\end{equation}
and
\begin{equation}\label{relmom1}
k' = \frac{k - (1+e)(k\cdot n)n}{\sqrt{1 + (e^2-1)(k\cdot n)^2}}\
.
\end{equation}

We now have what we need to write down the gain term in the
collision kernel in weak form using the inelastic reflection map:
\begin{equation}\label{weakgainref}
\langle Q_e^+(f,f), \varphi\rangle = \frac{1}{4\pi}
\int_{\R^3}\int_{\R^3}\int_{S^2} f(v) f(w)\varphi(v') \widetilde
\Phi(|u|)\widetilde B(k\cdot n) \dd n \dd v \dd w
\end{equation}
where $\varphi$ and $k$ are as before, and again, $\widetilde
\Phi$ and $\widetilde B$ gives the relative rates of the various
kinematically possible collisions, and $\dd n$ is the uniform
measure on $S^2$ with total mass $4\pi$.

All formulas involving the Fourier transform, of which we shall
make extensive use, are simplest in the
$\sigma$--parameterization, and so we must translate the above
inelastic parameterization into  terms of $\sigma$. This is easy
to do using (\ref{rel}) and (\ref{rel2}). For example, the first
line of (\ref{inreflect}) can be written as
$$
v' = \frac{v+w}{2} + \frac{u}{2} - \frac{1+e}{2}(u\cdot n)n =
\frac{v+w}{2} + \frac{1-e}{4}u + \frac{1+e}{4}|u|\sigma\ ,
$$
where (\ref{rel}) was used to eliminate $n$ in favor of $\sigma$.
In the same way, we translate the second line, and find $ w' =
\frac{v+w}{2} - \frac{1-e}{4}u - \frac{1+e}{4}|u|\sigma$, and it
follows that $ u'  =  \frac{1-e}{2}u +  \frac{1+e}{2}|u|\sigma$,
so that
\begin{equation}\label{energy2s}
|u'|^2 = |u|^2\left(\frac{1+e^2}{2} +
\frac{1-e^2}{2}k\cdot\sigma\right)\ ,
\end{equation}
and
\begin{equation}\label{relmom2s}
k'  = \frac{(1-e)k + (1+e)\sigma}{\sqrt{2(1+e^2) + 2(1-e^2)k\cdot
\sigma}} \ .
\end{equation}
Finally, from (\ref{inreflect}) we have $(n\cdot u')n = -e (n\cdot
u)n$. Then from (\ref{rel}) we have
$$
u' - |u'|\sigma' = 2(u'\cdot n)n = -2e (u\cdot n)n =  -e(u - |u|\sigma)\ .
$$
solving for $\sigma'$ we find
$$
\sigma' = k' + e\frac{|u|}{|u'|}(k-\sigma)\ .
$$
Using the expressions just derived for $|u'|$ and $k'$, we find
\begin{equation}\label{relmom3s}
\sigma'  = \frac{(1+e)k + (1-e)\sigma}{\sqrt{2(1+e^2) +
2(1-e^2)k\cdot \sigma}}\ .
\end{equation}

This gives us the {\em inelastic swapping map with restitution
coefficient} $e$, denoted $C_{s,e}$: $C_{s,e}(v,w,\sigma) =
(v',w',\sigma')$ where
\begin{equation}\label{inq}
\begin{cases} v'  &= {\displaystyle \frac{v+w}{2} + \frac{1-e}{4}(v-w)
-\frac{1+e}{4}|v-w|\sigma}\\[3mm]
w'  &= {\displaystyle \frac{v+w}{2} - \frac{1-e}{4}(v-w) +
\frac{1+e}{4}|v-w|\sigma}\\[3mm]
\sigma'  &= {\displaystyle \frac{(1+e)k +
(1-e)\sigma}{\sqrt{2(1+e^2) + 2(1-e^2)k\cdot \sigma}}}
\end{cases}
\end{equation}

We now have what we need to write down the gain term in the
collision kernel in weak form for inelastic collision in the
$\sigma$ representation: The basic expression that defines the
gain term is
\begin{equation}\label{wgs3}
\langle Q_e^+, \varphi\rangle = \frac{1}{4\pi}
\int_{\R^3}\int_{\R^3}\int_{S^2} f(v) f(w)\varphi(v') \Phi(|u|)
B(k\cdot \sigma) \dd\sigma \dd v \dd w
\end{equation}
where $\varphi$ is any test function, and as above $k$ denotes the
unit vector in the direction of $v-w$, and $\Phi$ and $B$ gives
the relative rates of the various kinematically possible
collisions, and $\dd \sigma$ is the uniform measure on $S^2$ with
total mass $4\pi$. Let us remark that being the $n$-representation
more physically meaningful, the $\sigma$-representation is a nice
mathematical device to derive estimates in both the weak and the
strong formulation.

\subsection{Strong form of the inelastic collision kernel}

Finally, we want to derive the strong form of the gain term, and
for this we need the {\em precollisional velocities} $v^*$ and
$w^*$. That is, given a collision map $C$, we define
$(v^*,w^*,\omega^*) = C^{-1}(v,w,\omega)$. It is very easy to
invert the the transformation  $(v,w,n) \mapsto (v',w',n')$ in the
reflection parameterization (\ref{inreflect}): Simply use a
restitution coefficient of $1/e$.  This gives us
$C_{r,e}^{-1}(v,w,n) =  (v^*,w^*,n^*)$ where
\begin{equation}\label{inreflectinv}
\begin{cases}  v^*  &=\  v - {\displaystyle\frac{1+e}{2e}((v-w)\cdot n)
n}\\[2mm]
 w^*  &=\   w +  {\displaystyle \frac{1+e}{2e}((v-w)\cdot n)
 n}\\[2mm]
 n^*  &=\   n
\end{cases}
\end{equation}
Note that with $u^* = v^*-w^*$, we have
\begin{equation}\label{energy2}
|u^*|^2 = |u|^2 + \left(\frac{1}{e^2}-1\right)(u\cdot n)^2   =
|u|^2 \left( 1 + \left(\frac{1}{e^2} -1\right)(k\cdot n)^2\right)\
\end{equation}
and $u^*\cdot n = -\frac{1}{e}u\cdot n$. Combining these we have
\begin{equation}\label{energy2c}
k^*\cdot n = -\frac{k\cdot n}{\sqrt{e^2 + (1-e^2)(k\cdot n)^2 }}\
.
\end{equation}
Now, the determinant of the Jacobian matrix for the transformation
in  (\ref{inreflect}) is easily seen to be
$$
\left(\frac{1-e}{2}\right)^2 -  \left(\frac{1+e}{2}\right)^2 = -e\ ,
$$
independent of $n$, so that the Jacobian itself is $e$. Thus, the
determinant of the Jacobian matrix of the inverse transformation
is $-1/e$, also independent of $n$, so the Jacobian itself is
$1/e$. We thus can rewrite (\ref{weakgainref}) as
\begin{equation}\label{weakgainref2}
\langle Q_e^+, \varphi\rangle = \frac{1}{4\pi}
\int_{\R^3}\int_{\R^3}\int_{S^2} f(v^*) f(w^*)\varphi(v)
\widetilde\Phi (|u^*|) \frac{\widetilde B(k^*\cdot n)}{e} \dd n
\dd v \dd w
\end{equation}
where $\varphi$ and $\widetilde B$ are  as before, and $k^* = v^*
- w^*$. Taking into account \eqref{energy2} and (\ref{energy2c})
and the fact that $\widetilde B$ is even,
\begin{equation}\label{sgalt1}
Q_e^+(f,f)(v)   = \frac{1}{4\pi} \int_{\R^3}\int_{S^2} f(v^*)
f(w^*) \widetilde \Phi_e^+(|u|,k\cdot n) \widetilde B_e^+
\left(k\cdot n \right)\dd n \dd w\ ,
\end{equation}
where $\widetilde B_e^+$
\begin{equation}\label{betildef}
\widetilde B_e^+(s) = \widetilde B\left(\frac{s}{\sqrt{e^2
+(1-e^2)s^2 }}\right)\frac{1}{e} \ ,
\end{equation}
and
\begin{equation}\label{phietildef}
\widetilde \Phi_e^+(r,s) = \widetilde\Phi\left(\frac{r}{e}
\sqrt{e^2 +(1-e^2)s^2}\right) \ .
\end{equation}

Then, using the identities in (\ref{rel}) once more, we may
translate (\ref{inreflectinv}) into the $\sigma$ parameterization,
and obtain $C_{s,e}^{-1}(v,w,\sigma) =  (v^*,w^*,\sigma^*)$ where
\begin{equation}\label{inqnv}
\begin{cases}  v^*  &=\  {\displaystyle \frac{v+w}{2} - \frac{1-e}{4e}(v-w) +
\frac{1+e}{4e}|v-w|\sigma}\\[3mm]
 w^*  &=\     {\displaystyle\frac{v+w}{2} + \frac{1-e}{4e}(v-w) - \frac{1+e}{4e}|v-w|\sigma
 }\\[3mm]
\sigma^* &= {\displaystyle \frac{(1+e)k -
(1-e)\sigma}{\sqrt{2(1+e^2) - 2(1-e^2)k\cdot \sigma}}}
\end{cases}
\end{equation}
Along the way, we find, using (\ref{rel}) and (\ref{rel2}) as
before,
\begin{equation}\label{energy2c2}
|u^*|^2 = \frac{|u|^2}{2e^2} \left( (1+e^2) - (1-e^2)(k\cdot
\sigma) \right)\ .
\end{equation}
and
\begin{equation}\label{relmom5s}
k^*  = \frac{(1+e)\sigma - (1-e)k}{\sqrt{2(1+e^2) - 2(1-e^2)k\cdot
\sigma}} \ .
\end{equation}
Combining these we have
\begin{equation}\label{energy2cc}
|k^*\cdot n| = \sqrt{\frac{1-k\cdot\sigma}{(1+e^2) -
(1-e^2)k\cdot\sigma}}\ .
\end{equation}

Now, we use (\ref{rel4}) and (\ref{energy2c2}) to translate
$\widetilde B(k^*\cdot n)\dd n$ into terms of $\sigma$. We find,
remembering that $\widetilde B$ is even:
$$
\widetilde B(k^*\cdot n)\dd n = \frac{1}{4} \widetilde B\left( \sqrt{\frac{1-k\cdot\sigma}
{(1+e^2) -
(1-e^2)k\cdot\sigma}}\right)\sqrt{\frac{2}{1-k\cdot\sigma}}\dd\sigma
\ .
$$
Next, we use (\ref{change3}) to express this in terms of $B$ in
place of $\widetilde B$: We obtain:
$$
\widetilde B(k^*\cdot n)\dd n =
B\left(\frac{(1+e^2)k\cdot \sigma - (1-e^2)}{(1+e^2) -
(1-e^2)k\cdot\sigma}\right) \frac{\sqrt{2}}{\sqrt{{(1+e^2) -
(1-e^2)k\cdot\sigma}}}\dd \sigma\ .
$$
Moreover, using \eqref{energy2c}, we have
$$
\widetilde \Phi(|u^*|)=\Phi(|u^*|)=\Phi\left(\frac{|u|}{\sqrt{2}e}
\sqrt{(1+e^2) - (1-e^2)(k\cdot \sigma)}\right)\ .
$$
Therefore, defining the function $B_e^+$ by
$$
B_e^+(s) = B\left(\frac{(1+e^2)s - (1-e^2)}{(1+e^2) -
(1-e^2)s}\right) \frac{\sqrt{2}}{\sqrt{{(1+e^2) -
(1-e^2)s}}}\frac{1}{e}
$$
and the function
$$
\Phi_e^+(r,s)=\Phi\left(\frac{r}{\sqrt{2}e} \sqrt{(1+e^2) -
(1-e^2)s}\right)\ ,
$$
we have
\begin{equation}\label{sgalt2}
Q_e^+(f,f)(v)   = \frac{1}{4\pi} \int_{\R^3}\int_{S^2} f(v^*)
f(w^*) \Phi_e^+(|u|, k\cdot \sigma) B_e^+ (k\cdot \sigma)\dd
\sigma \dd w \ .
\end{equation}

The change of variable formulas that we have deduced in this
section, suffice not only to allow us to write down the strong
form of the collision kernel, but also to prove the following
somewhat more general result:

\begin{thm}\label{backforth}
Let $K$ be any continuous real valued function on $\Xi \times
\Xi$. Then
\begin{multline}\label{bfs}
\frac{1}{4\pi}\int_{\R^3}\int_{\R^3} \int_{S^2}
K[C_{s,e}^{-1}(v,w,\sigma),(v,w,\sigma)] \Phi_e^+(|u|, k\cdot
\sigma) B_e^+(k\cdot\sigma) \dd\sigma\dd w\dd v\\ =
\frac{1}{4\pi}\int_{\R^3}\int_{\R^3} \int_{S^2}
K[(v,w,\sigma),C_{s,e}(v,w,\sigma)] \Phi(|u|) B(k\cdot\sigma)
\dd\sigma\dd w\dd v
\end{multline}
and
\begin{multline}\label{bfr}
\frac{1}{4\pi}\int_{\R^3}\int_{\R^3} \int_{S^2}
K[C_{r,e}^{-1}(v,w,n),(v,w,n)] \widetilde \Phi_e^+(|u|, k\cdot n)
\widetilde B_e^+(k\cdot n) \dd n\dd w\dd v\\ =
\frac{1}{4\pi}\int_{\R^3}\int_{\R^3} \int_{S^2}
K[(v,w,n),C_{r,e}(v,w,n)] \widetilde \Phi(|u|) \widetilde B(k\cdot
n) \dd n\dd w\dd v\ ,
\end{multline}
where $\Phi=\widetilde \Phi$ and $B$ and $\widetilde B$ are
related by \eqref{change3}.
\end{thm}

%%%%%%%%%%%%%%%%%%%%%%%%%%%%%%%%%%%%%%%%%%%%%%%%%%%%%%%%%%%%%%%%%%

\section*{Acknowledgements} The authors are grateful to G. Toscani
for the comment regarding the new proof of the formulas in
\eqref{bonn4}-\eqref{bonn5} in the elastic case. This work was
begun when EC and MCC were visiting the Centre de Recerca
Matem\`atica (CRM) in Barcelona whom we thank for the hospitality.
We also acknowledge IPAM (UCLA) where this work was finished.

%%%%%%%%%%%%%%%%%%%%%%%%%%%%%%%%%%%%%%%%%%%%%%%%%%%%%%%%%%%%%%%%%%%%%%%%%

\end{document}